\documentclass[a4paper,11pt]{article}

\usepackage{amssymb,amsmath,amsfonts,placeins,pbox,multirow,mathtools,bbold}
\usepackage{graphicx,rotate,color,slashed,cite,epstopdf,verbatim,url,multirow,booktabs}
\usepackage{epstopdf}
\usepackage{subfig}
\usepackage{graphicx}
\usepackage{changepage}
\usepackage[normalem]{ulem}
\usepackage{cancel}
\graphicspath{{./Figures/}}
\usepackage{array}
\renewcommand{\arraystretch}{1.5} 
\usepackage[a4paper, left=2cm, right=2cm, top=2cm, bottom=2cm]{geometry}

\usepackage[colorlinks=true,
            linkcolor=magenta,
            urlcolor=blue,
            citecolor=blue]{hyperref}
\usepackage[utf8x]{inputenc}

\usepackage[capitalise]{cleveref}
\usepackage[dvipsnames]{xcolor}

\crefname{section}{Sec.}{Secs.}
\crefname{table}{Tab.}{Tabs.}
\crefname{figure}{Fig.}{Figs.}
\crefname{equation}{Eq.}{Eqs.}
\crefname{appendix}{Appendix}{Appendix}

\numberwithin{equation}{section}

\def\ssbh#1//#2//{\ensuremath{\xrightarrow [\substack{#2}]
    {\parbox{3cm}{\hfil $\scriptstyle \langle #1 \rangle$ \hfil}}}}

\let\CapTion=\caption
\def\caption#1{\CapTion{\em #1}}

\title{\bf Vector-like tops from first generation quarks:\\the role of width and coupling chiralities\\ in same-charge production at the LHC}

\author{Stefano Moretti$^{a,b}$\thanks{stefano.moretti@physics.uu.se;~s.moretti@soton.ac.uk}\,, Luca Panizzi$^{c,d}$\thanks{luca.panizzi@unical.it}\, and Liangliang Shang$^{a,e}$\thanks{283475978@qq.com}
\\ \medskip 
${}^{a}$\small\em Department of Physics and Astronomy, Uppsala University, Box 516, SE-751 20 Uppsala, Sweden\\[-14pt]
${}^{b}$\small\em School of Physics and Astronomy, University of Southampton, Highfield, Southampton SO17 1BJ, UK\\[-8pt]
{}$^{c}$\small\em Dipartimento di Fisica, Universit\`a della Calabria, I-87036 Arcavacata di Rende, Cosenza, Italy\\[-8pt]
{}$^{d}$\small\em INFN-Cosenza, I-87036 Arcavacata di Rende, Cosenza, Italy.\\[-8pt]
{}$^{e}$\small\em School of Physics, Henan Normal University, Xinxiang 453007, China
}
\date{}

\begin{document}

\maketitle


\begin{abstract}
In this paper, we tension the scope of the $q\bar q, gg\to T \bar T$ production process against that of the same-charge $uu\to TT$ one in 
searching for Vector-Like Quarks (VLQs) with electromagnetic charge $+2/3$ at the Large Hadron Collider (LHC). The study is conducted by considering a simplified model in which such new states couple exclusively to Standard Model quarks of the first generation through $W^\pm, Z$ and $H$ currents. We consider both left-handed or right-handed dominant chiralities, representative of $SU(2)_L$ singlets or doublets, and the total widths are allowed to reach large values relatively to the VLQ masses. By deconstructing the signal into all its components, we assess current and projected constraints, emphasize the role of interference terms and isolate peculiar features which can be used for dedicated searches at the LHC or future colliders.
\end{abstract}

\bigskip

\newpage
\tableofcontents
\bigskip
\hrule
\bigskip
\section{Introduction}
\label{intro}

Vector-Like Quark (VLQ) searches have become a cornerstone of analyses at the Large Hadron Collider (LHC), having been investigated in numerous experimental studies.
In fact, particular attention has historically been given to these 
particle states in the case of a VLQ  with an Electro-Magnetic (EM) charge of $+2/3$, denoted as $T$, analogous to the top quark in the Standard Model (SM). This focus is unsurprising, as such a "top companion" can play a significant role in fundamental theories of the Electro-Weak (EW) scale (see \cite{Branco:2022gja} for a concise perspective on their role). 
These (presumed) new states of Nature are nothing but spin-1/2 coloured fermions with Left-Handed (LH) and Right-Handed (RH) components transforming in the same way under the SM gauge group $SU(3)_C \times SU (2)_L \times U (1)_Y$, thus unlike the so-called `chiral’ quarks of the SM, where only the LH component intervenes. Such a property makes VLQs fundamentally different from their SM counterparts, though, as it implies that they do not need to acquire their mass via the Higgs mechanism.

However, due to their colour charge, VLQs can  abundantly be  produced at the LHC, provided their mass lies within the TeV range. To date, a large part of systematic studies have primarily focused on the pair production of oppositely charged VLQs, {\it i.e.}, the channel $q\bar{q},gg\to T\bar{T}$, largely because of its substantial cross-section driven by Quantum Chromo-Dynamics (QCD) interactions. Furthermore, a common assumption in this approach is that $T$ VLQs couple exclusively to the third-generation quarks of the SM, namely, the $t$- and $b$-quarks.

Here, we instead address the case of pair production of VLQs with identical EM charge $+2/3$ (and the charge-conjugate case with EM charge $-2/3$), induced by interactions between $T$ VLQs and SM (anti)quarks of the first generation. 
These types of interactions do not easily arise in the theoretical frameworks commonly invoked to justify the potential existence of VLQs. For example, in composite Higgs models with partial compositeness, VLQs mix with SM quarks through Yukawa interactions involving the Higgs boson, leading to significant interactions only when the VLQ mixes with the heaviest quark, namely, the top quark. However, other theoretical frameworks predict VLQs that interact with lighter generations. Examples include the Kaluza-Klein even partners of SM quarks in universal extra-dimensional scenarios~\cite{Cacciapaglia:2009pa} or models that aim to explain the mass hierarchies among SM fermions~\cite{Jana:2021tlx,Davighi:2022bqf}.

There are also compelling phenomenological arguments to study same-charge $T$ VLQ pair production at the LHC. On the one hand, despite being an EW process, the cross-section for $uu\to TT$ production can be significant due to the valence nature of $u$-quarks: this provides an enhancement in their Parton Distribution Functions (PDFs) compared to gluons, particularly at the large $x$ values required for producing final states with substantial rest mass (on the order of TeV). On the other hand, the decay patterns arising from $TT$ (and to a lesser extent, $\bar{T}\bar{T}$) decays are notably distinct from those of $T\bar{T}$ production: this motivates new search strategies for $TT$ production and decay, as the corresponding signal is subject to reduced SM background compared to the QCD-dominated $T\bar{T}$ channel.

These aspects have also been highlighted in~\cite{Buckley:2020wzk,Cui:2022hjg}, where same-sign pair production was considered within the Narrow Width Approximation (NWA). However, an additional challenge arises when VLQ production processes depend on the same couplings that also govern their decays. When such couplings are large, not only does the cross-section increase, but the VLQ width also becomes significant. This necessitates departing from the typical NWA used in most analyses and requires the inclusion of interference effects. This affects the kinematical properties of the objects in the final state and may lead to experimentally visible differences if the deviations are larger than detector resolutions.

The purpose of this paper is to study in detail the LHC phenomenology of the $uu \to TT$ (and its charge-conjugate) process using a simplified model approach. This approach embeds $T$ VLQs within phenomenological scenarios defined by fixed relations between Branching Ratios (BRs) and incorporates VLQs in real singlet and doublet representations, where the mixing angle determines the couplings, partial widths and BRs themselves. We will focus exclusively on interactions of $T$ (and $\bar{T}$) states with $u$- and $d$-(anti)quarks mediated by the SM EW gauge bosons, $W^\pm$ and $Z$, and the Higgs boson $H$. Our analysis will systematically include large width and interference effects.

In the latter, we will employ a deconstructive strategy to better understand the role of individual subprocesses in shaping the kinematic features of the final state. This approach will help identify the signal regions that are most sensitive to specific scenarios.
Through this strategy, we will delineate the accessible parameter space, expressed in terms of the relevant masses and couplings of the new fermionic states, as well as their mixing angles with SM quarks. This analysis will take into account both theoretical constraints and experimental limits, either directly or by recasting existing results from the LHC and other facilities, with particular attention given to Atomic Parity Violation (APV) data in the context of real singlet and doublet VLQ representations.
We also provide a comprehensive discussion of the prospects for observing this production channel at the LHC across various final states in the near future.

The structure of the paper is as follows. In the next section, we introduce our model framework and its parameterisation. This is followed by a discussion of key technical aspects of our analysis, leading to the presentation of our results. Finally, we provide a summary and some conclusions.

\section{Simplified model and parametrisation of the processes}
\label{sec:SMresults}

As intimated, we will consider for this analysis a simplified model where we extend the SM with one VLQ with charge $+2/3$ (and it charge-conjugate state) and allowed it to interact with the SM first generation quarks as well as $W$, $Z$ and $H$ bosons. The interaction Lagrangian in the mass eigenstates can be written as
\begin{equation}
\mathcal{L}_{\rm int} = \frac{g_W}{\sqrt{2}} \left(\kappa^W_L \bar{T} \slashed W^+ P_L d  +  L\leftrightarrow R\right) + \frac{g_W}{2 c_W} \left(\kappa^Z_L \bar{T} \slashed Z P_L u +  L\leftrightarrow R\right) + H \left(\kappa^H_L \bar{T} P_L u +  L\leftrightarrow R\right) + {\rm h.c.}\;,
\label{eq:LagTSM}
\end{equation}
where $g_W$ is the coupling constant of ${SU(2)}_L$, with $c_w =\cos \theta_W$, where $\theta_W$ is the weak mixing angle. The three dimensionless couplings $\kappa^{W,Z,H}_{L,R}$ are in principle free parameters, which depend on the specific underlying theory of new physics containing the VLQ. However, it is known that, depending on the representation of the VLQ under the EW gauge group $SU(2)_L\times U(1)_Y$, one of the two chirality projections is always suppressed with respect to the other by a factor of order $m_q/m_T$~\cite{delAguila:2000rc,Buchkremer:2013bha}, where $m_q$ is the SM quark mass and $m_T$ is the $T$ mass. Therefore, in the following, we will always set the coupling of one of the two chiralities to zero and consider only purely LH or purely RH couplings. We will discuss the kinematical differences associated to either choice in \cref{sec:features}. 

Due to the freedom in choosing the couplings with gauge bosons, this Lagrangian is not gauge invariant {\it per se}, but this is not a concern for this analysis because what matters is that it can be mapped to represent classes of fully consistent theoeretical models which contain these kind of interactions. It can be derived, for example, by starting from any number of different representations of the flavour eigenstates of VLQs in the unbroken EW phase, allowing them to interact with the first generation quarks through new Yukawa parameters. The mass eigenstates can then be obtained by diagonalising the mixing matrix, leading to the general structure of above Lagrangian~\cite{Buchkremer:2013bha,Fuks:2016ftf}, and the lightest one can be considered for performing phenomenological analyses. 

\begin{figure}[h!]
\centering
\includegraphics[width=.7\textwidth]{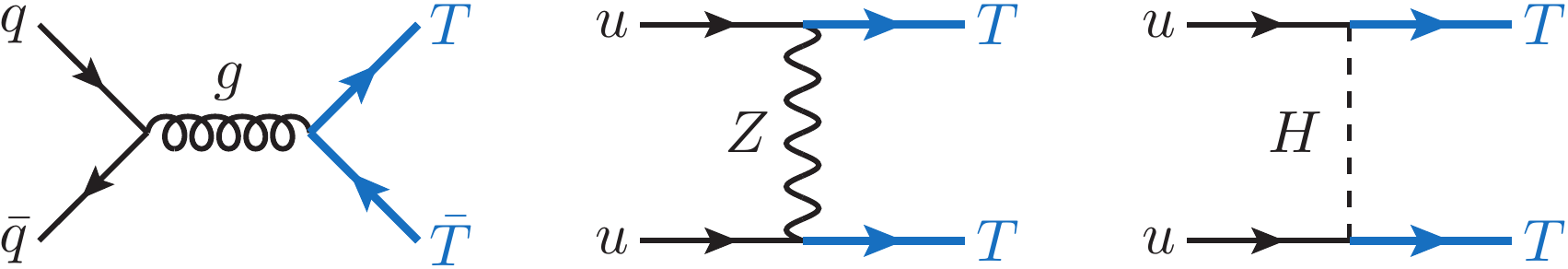}
\caption{\label{fig:topologies} Representative Feynman diagrams for QCD $T\bar T$ production ({left}) and $Z$ and $H$ ({center and right, respectively}) mediated $TT$ production.}
\end{figure}

Let's start by considering final states arising from processes in which the $T$ is produced in pairs. Unlike in scenarios where VLQs interact exclusively with SM quarks of the third generation, the interactions to up and down quarks lead to the possibility of producing two $T$ states with the same charge via $t,u$-channel exchange of $Z$ or $H$ bosons~\cite{Buchkremer:2013bha}. Representative Feynman diagrams associated with such processes are shown in \cref{fig:topologies}.

The same-charge $TT$ production processes are enhanced by the large $u$-quark PDFs at the LHC and, therefore, if the interaction strengths in the Lagrangian~\cref{eq:LagTSM} are large enough, these processes can lead to high cross-sections, potentially competitive with or even larger than the QCD ones for $T\bar T$ production. The other $t,u$-channel processes $T\bar T$ and $\bar T \bar T$ are subdominant because initiated by at most one valence quark PDF. 

According to the Lagrangian in \eqref{eq:LagTSM}, the $T$ VLQ has three possible decay channels: $Wd$, $Zu$ and $Hu$. Depending on the relative BRs and on the subsequent decays of the SM bosons, the phenomenological consequences are multiple.
\begin{itemize}
\item If the $T$ does not decay into $W$ bosons with subsequent leptonic decays, it is not possible to distinguish $T\bar T$ from $TT$ production through particle selections, but only -- in principle -- through the cross-section and the different kinematical properties of the objects in the final state. 
\item In case of significant decay rates associated to $T$ decays into $Wd$ with $W^+\to l^+\nu$ subsequent decays, the EW mediated $t,u$-channel processes feature a same-charge di-lepton final state, which can be exploited to isolate the signal from the background.
\end{itemize}

\begin{figure}[h!]
\centering
\includegraphics[width=.7\textwidth]{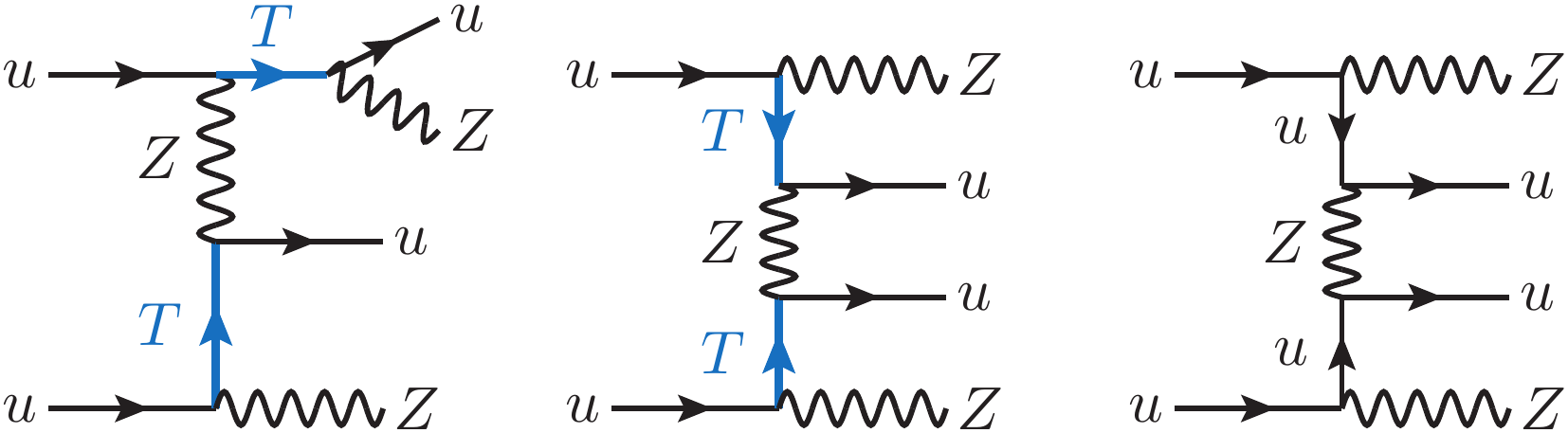}
\caption{\label{fig:LWtopologies} Representative topologies for the $ZuZu$ final state which can be relevant if the $T$ width is not narrow, featuring only one resonant ({left}) or no resonant ({centre}) $T$ propagators as well as a SM background process which also interferes ({right}).}
\end{figure}

To achieve a significant cross-section for $TT$ production, couplings must not be too small: this in turn implies that the $T$ total width can be relatively large with respect to its mass. In this case, an approach based on the NWA would not describe the final state kinematics with sufficient accuracy. The processes will be thus analysed without imposing resonant propagation of the $T$'s, {\it i.e.}, by simulating the full $2\to 4$ processes of type $pp\to qqBB$ or $pp\to q\bar{q}BB$  where $q$ and $B$ are SM quarks and bosons (gauge or Higgs) of either charge combination, respectively. For a fixed 4-body final state, the only two parameters which determine the kinematical properties of the final state objects are the $T$ mass, $m_T$, and total width, $\Gamma_T$. The couplings at both ends of the $T$ propagators can be factorised and rescaled to increase or decrease the cross-section without affecting the kinematical distributions. Following a deconstructive approach already used in previous studies \cite{Deandrea:2021vje,Moretti:2023dlx,Banerjee:2024zvg}, the cross-sections of the $2\to4$ processes can be written as
\begin{subequations}
\begin{align}
\sigma_{T\bar T}(m_T,\Gamma_T)&=
\sum_{a,\bar b} \kappa_{a}^2\kappa_{b}^2 ~\hat\sigma_{pp\to a \bar b}(m_T,\Gamma_T) + \sum_{a,\bar b} \kappa_{a}\kappa_{b} ~\hat\sigma_{pp\to a \bar b}^{int}(m_T,\Gamma_T)\;,\\
\sigma_{TT}(m_T,\Gamma_T)&=
\sum_{a,b} \kappa_{Z}^4\kappa_{a}^2\kappa_{b}^2 ~\hat\sigma^Z_{pp\to ab}(m_T,\Gamma_T)+\sum_{a,b} \kappa_{H}^4\kappa_{a}^2\kappa_{b}^2 ~\hat\sigma^H_{pp\to ab}(m_T,\Gamma_T)\nonumber\\
& + \sum_{a,b} \kappa_{Z}^2\kappa_{a}\kappa_{b} ~\hat\sigma^{ZB_{\rm int}}_{pp\to ab}(m_T,\Gamma_T) + 
\sum_{a,b} \kappa_{H}^2\kappa_{a}\kappa_{b} ~\hat\sigma^{HB_{\rm int}}_{pp\to ab}(m_T,\Gamma_T)\nonumber\\
& + \sum_{a,b} \kappa_{Z}^2\kappa_{H}^2\kappa_{a}^2\kappa_{b}^2 ~\hat\sigma^{ZH_{\rm int}}_{pp\to ab}(m_T,\Gamma_T)\;,
\end{align}
\label{eq:sigmahats}
\end{subequations}

\noindent where $\{a,b\}=\{W^+d, Zu, Hu\}$ and $\{\bar a, \bar b\}$ are their charge-conjugates. The interference contributions between the $Z$ and $H$ mediated topologies and the SM background and between themselves have been included. They can become relevant because of the presence of topologies where there is only one on-shell $T$ or where the VLQ propagates only in $t,u$-channels, as shown in \cref{fig:LWtopologies}, as such diagrams belong to the same gauge invariant set involving the case of $T\to Zu$ decays in the $uu\to TT$ process. Furthermore, there are interferences with SM topologies too, like the last diagram in \cref{fig:LWtopologies}. Finally, none of the contributions due to these diagrams can be factorised in the NWA simultaneously to the $uu\to TT\to ZuZu$ process.

\section{Technical aspects of the analysis}
\label{sec:tools}

We have implemented the simplified model of Eq.~\eqref{eq:LagTSM} in a dedicated {\tt UFO}~\cite{Degrande:2011ua, Darme:2023jdn} model derived from the original {\tt eVLQ} model described in~\cite{Banerjee:2022xmu}, to account for the technical requirements needed to isolate the different contributions in Eq.~\eqref{eq:sigmahats}: these  amount to defining specific coupling orders for each interaction of the $T$ state, 
using a procedure analogous to those described in~\cite{Deandrea:2021vje,Moretti:2023dlx}. We use a standard tool chain: {\sc MG5\_aMC}~\cite{Alwall:2014hca} for the matrix-element part and {\tt Pythia 8}~\cite{Sjostrand:2014zea} for parton shower and hadronisation. Due to the exploratory nature of this analysis, the simulations have been performed at Leading Order (LO). The LO set of the {\sc NNPDF 4.0} PDFs~\cite{NNPDF:2021njg} has been used. 
For a uniform evaluation of pure signal and interference terms, all parton level simulations have been performed imposing standard minimal cuts ($p_T(j)>20$ GeV, $|\eta(j)|<5$ and $\Delta R(jj)>0.4$) and the renormalisation and factorisation scales have both been fixed to half the transverse mass of the final state particles, on an event-by-event basis. 

We have evaluated LHC constraints on the new physics signal by recasting experimental data through the {\tt MadAnalysis 5} framework~\cite{Conte:2012fm, Conte:2014zja, Conte:2018vmg}, which internally uses the anti-$k_T$ algorithm~\cite{Cacciari:2008gp} implemented in {\sc FastJet}~\cite{Cacciari:2011ma} as well as the detector simulations of {\tt Delphes~3}~\cite{deFavereau:2013fsa} and {\tt SFS}~\cite{Araz:2020lnp}. Considering that the final states have large jet multiplicity but also potential same-charge di-leptons (depending on the interactions of the $T$ VLQ), we have tested our signal against various recasts of ATLAS~\cite{ATLAS:2020syg,ATLAS:2019lsy,ATLAS:2019wgx,ATLAS:2019lff,ATLAS:2021kxv} and CMS~\cite{CMS:2018ikp,CMS:2019lwf,CMS:2019zmd} searches, available in the {\tt MadAnalysis 5} public analysis database\footnote{A more recent ATLAS search targets pair production of $T$ VLQs decaying to $Wd$, reinterpreting the results also for $Zu$ and $Hu$ decays~\cite{ATLAS:2024zlo}. While very relevant for our purposes, a dedicated recast and signal testing will be left for a future update of our results. We will nevertheless interpret the latter in the context of the bounds from this search in the final discussion.}. 

We have scanned the $\{m_T,\Gamma_T/m_T\}$ parameter space in the region $m_T=\{130, 150, 200, 400, \dots, \- 4000\}$ GeV and $\Gamma_T/m_T=\{0.001,0.002,0.005,\dots,0.5\}$ (in logarithmic scale).
From the recast results we have extracted the experimental efficiencies $\epsilon_i$ (which are function only of $m_T$ and $\Gamma_T$) associated with each SR of each search and each $\sigma_i$ term of \cref{eq:sigmahats} ($i$ runs over all the subprocesses) and used these together with the integrated luminosities of each search to compute the number of signal events for each point of the scan according to the general expression 
\begin{equation}
S(m_T,\Gamma_T,{\rm SR})= \mathcal L({\rm SR})\sum_i \sigma_i(m_T,\Gamma_T) \epsilon_i(m_T,\Gamma_T, {\rm SR})\;.
\end{equation}
The $k$ couplings of \cref{eq:sigmahats}, necessary to obtain the $\sigma_i$ terms, are computed analytically for each benchmark (defined in the next section) knowing that they have to reconstruct the correct total width and BRs. We have then used this together with the information about background and uncertainties from the experimental data to compute the significance through the  method~\cite{Read:2002hq} and select the SR with highest exclusion limit. 

Finally, we summarise the technical limitations of our analysis. For all final states, we have considered contributions which can be reduced to the NWA pair production ones in the limit of small couplings. These are contributions which contain an even number of $T$ propagators in the amplitude squared (either resonant or not); there can be contributions involving the propagation of only one or three VLQ leading to the considered final states, but in the limit of small coupling they reduce to pure (and kinematically suppressed) interference processes between resonant single production with emission of a further jet, and resonant pair production, and thus have been ignored. Furthermore, all interference terms with the SM background have been computed assuming the same orders in $\alpha_s$ of QCD for signal and SM amplitudes. All these limitations will be removed in a forthcoming analysis addressing final states with different object multiplicities.

\section{Numerical results}
\label{sec:results}

We will provide constraints for different phenomenological benchmarks, usually considered in this kind of analyses and related to the proportion of BRs of $T$ into $Wd$, $Zu$ and $Hu$, namely, when $T$ exclusively decays in either of the three, as well as when $T$ has singlet-like or doublet-like BRs relations, ${\rm BR}_{T\to Wd}:{\rm BR}_{T\to Zu}:{\rm BR}_{T\to Hu}=2:1:1$ and ${\rm BR}_{T\to Wd}:{\rm BR}_{T\to Zu}:{\rm BR}_{T\to Hu}=0:1:1,$ respectively. These benchmarks are `phenomenological', in the sense that these exact relations between BRs cannot be achieved in a scenario where $T$ is the only new particle beyond the SM and belonging to a definite representation under the EW gauge group (the singlet-like and doublet-like ones can only be reached asymptotically for large VLQ masses). Further new physics must be implied to achieve these relations, in the form of heavier VLQs or extended scalar sectors, which modify the mixing matrix patterns. For completeness, therefore, we are going to also consider the cases in which $T$ is truly the only new Beyond the SM (BSM) particle and belongs to a singlet or is part of a $(T~B)$ doublet representation (where, for simplicity, we assume that the $B$ VLQ does not mix with the SM bottom quark) under the EW gauge group. The relations between BRs are therefore determined by the $T-t$ mixing angle, which in turn also determines the total width of the VLQ via the mathematical constraint that sines and cosines of the mixing angle must be between 0 and 1. For these scenarios, we will consider only the case with dominant chirality, LH for doublet and RH for singlet. A summary of the benchmark scenarios considered in this analysis is provided in \cref{tab:benchmarks}.
\begin{table}[h!]
\centering
{\renewcommand{\arraystretch}{1}
\begin{tabular}{ccccc}
\toprule
Benchmark description & ${\rm BR}_{T\to Wd}$ & ${\rm BR}_{T\to Zu}$ & ${\rm BR}_{T\to Hu}$ & chiralities \\
\midrule
1:0:0 & 1 & 0 & 0 & LH/RH \\
\midrule
0:1:0 & 0 & 1 & 0 & LH/RH \\
\midrule
0:0:1 & 0 & 0 & 1 & LH/RH \\
\midrule
2:1:1 (singlet-like) & 0.5 & 0.25 & 0.25 & LH/RH \\
\midrule
0:1:1 (doublet-like) & 0 & 0.5 & 0.5 & LH/RH \\
\midrule
\multirow{2}{*}{true $T$ singlet} & \multicolumn{3}{c}{function of $m_T$ and $\Gamma_T$} & \multirow{2}{*}{only LH} \\
& \multicolumn{3}{c}{($\sim$ 2:1:1 for large $m_T$)} \\
\midrule
true $(T~B)$ doublet & \multicolumn{3}{c}{function of $m_T$ and $\Gamma_T$}   & \multirow{2}{*}{only RH}\\
(with no $B-b$ mixing)& \multicolumn{3}{c}{($\sim$ 0:1:1 for large $m_T$)} \\
\bottomrule
\end{tabular}
}
\caption{\label{tab:benchmarks} Summary of the benchmark scenarios used in this analysis.}
\end{table}

These scenarios can be strongly constrained by low energy observables or EW precision tests~\cite{Okada:2012gy,Chen:2017hak,Erdelyi:2024sls}, however, such constraints are usually sensitive to all BSM contributions from new particles propagating in loops and can receive cancellations from opposite effects once more complete theories are considered. For this reason, we will apply low energy constraints only to the scenarios where the VLQ is the only new state: if the VLQ interacts exclusively with SM first generation quarks, the most relevant constraint comes from deviations in APV observables~(see \cite{Arcadi:2019uif} for a review).\\

\begin{figure}[h!]
\centering
\includegraphics[width=.33\textwidth]{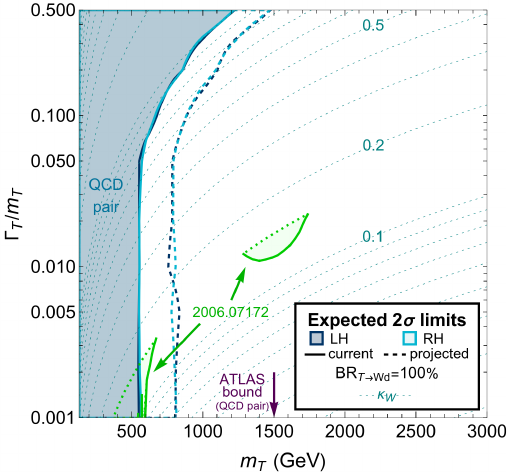}\hfill
\includegraphics[width=.33\textwidth]{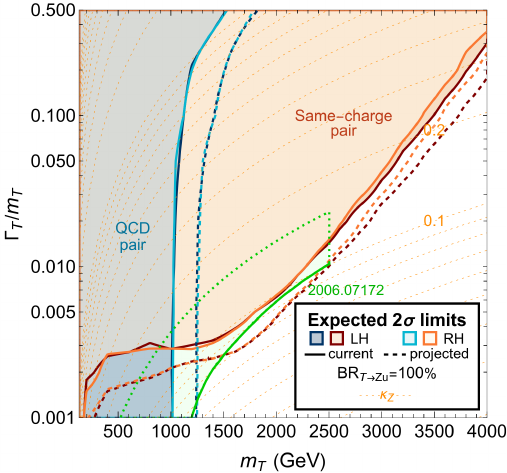}\hfill
\includegraphics[width=.33\textwidth]{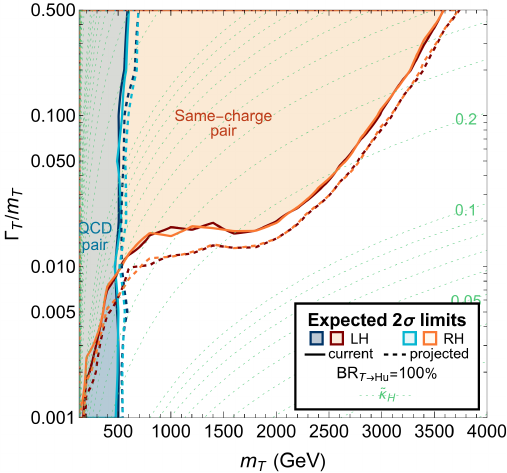}
\caption{\label{fig:limitsmaximalBRs} Comparison of constraints in the $\{m_T,\Gamma_T/m_T\}$ plane for $T\bar T$ pair production (excluded area in light blue) and $TT$ pair production (excluded area in light red) from a recast of multiple ATLAS~\cite{ATLAS:2020syg,ATLAS:2019lsy,ATLAS:2019wgx,ATLAS:2019lff,ATLAS:2021kxv} and CMS~\cite{CMS:2018ikp,CMS:2019lwf,CMS:2019zmd} searches. Projected $2\sigma$ bounds for the nominal Run 3 luminosity of 300/fb, assuming a 5\% systematics error for the background, are shown as dashed lines. From left to right, the $T$ BR is 100\% to $Wd$, $Zu$ and $Hu$, respectively. The values of the couplings from \cref{eq:LagTSM} are shown as dotted iso-lines in each panel; for the Higgs coupling we plot the quantity $\tilde\kappa_H\equiv (v/m_T) \kappa_H$, to show numerically when the asymptotic relation $\kappa\simeq\kappa_W\simeq\kappa_Z\simeq\tilde\kappa_H$ is obtained in the limit $v\ll m_T$. For the $Wd$ case, a comparison with the current most stringent bound from ATLAS~\cite{ATLAS:2024zlo} is shown for comparison (the same search has no sensitivity in the other cases). Bounds obtained in~\cite{Buckley:2020wzk} are also shown in green where applicable.}
\end{figure}

In \cref{fig:limitsmaximalBRs} we compare the LHC constraints for (opposite-charge) QCD $T\bar T$ pair production and (same-charge) $TT$ (plus charge-conjugate) pair production in the $\{m_T,\Gamma_T\}$ plane for the different maximal BR choices (100\% in either channel), considering both pure LH and pure RH couplings. 
In all cases, the QCD limit stays fairly independent of the width/mass ratio as long as the total width is small. The effect of finite width starts to be visible when the width is between 1\% and 5\% of the $T$ mass for the $Wd$ and $Zu$ cases, while for $Hu$ the dependence on the width is much milder. This is due to a combination of effects from the different scaling of the cross-sections and the different kinematics of the final state when considering VLQ interactions with scalars or vectors.

The dependence on the chirality of the coupling is noticeable only in the BR$_{T\to Wd}=100\%$ scenario, when the width is large. This is due to the increasing role of interference contributions with the SM background, for which only LH couplings are allowed due to the structure of EW interactions. The $Wd$ case does not allow for same-charge $TT$ pair production, as this requires non-zero couplings with $Zu$ or $Hu$. Nevertheless, this scenario allows for a comparison with the most recent ATLAS bound from~\cite{ATLAS:2024zlo}. Even if the ATLAS constraint applies only to a narrow width region, it is reasonable to assume that, if a recast is performed and signals with different $T$ widths are processed through it, the qualitative behaviour of the constraint as function of the width would not differ significantly from what obtained in our analysis. It is then interesting to notice, in this respect, that the excluded mass value obtained by our recast increases by about 700 GeV (1 TeV) between the NWA and the large width (50\% of the $T$ mass) limit with LH (RH) coupling assumption. 

For the $Zu$ and $Hu$ scenarios, constraints from same-charge $TT$ pair production become competitive with the QCD ones already for small values of the width/mass ratios and strongly dominate when the width is large, excluding very high values of the $T$ mass. In the $Zu$ case the QCD bound excludes masses below 1 TeV in the NWA region and increases to around 1.5 TeV when the width is 50\% of the mass. In fact, the $TT$ process starts to dominate when the width is larger than around 0.5\% of the mass and excludes a 4 TeV $T$ with width equal to 30\% of the mass. Even higher masses (with higher widths) would be excluded, but they are outside our simulation range. A very similar qualitative behaviour can also be seen in the $Hu$ case, where the $T\bar T$ bound of 500 GeV is overtaken by the one stemming from the $TT$ process when the width is 1\% of the mass, so that the exclusion reaches a 3.5 TeV mass limit for the highest width/mass value of our scan, which is 50\%.

It seems reasonable to assume that, if a new search is designed, which can be sensitive to these final states, an increase of the limits analogous to the $Wd$ final state would not change the qualitative picture which sees the $TT$ process being extremely relevant with respect to the $T\bar T$ one. For this purpose we show the projected expected limits (as dashed lines in the plots) for the nominal luminosity at the end of the LHC Run 3, corresponding to 300/fb\footnote{These projections have been obtained under the (optimistic) assumption that for all the considered searches the systematics uncertainty on the background becomes 5\%. Other assumptions are possible but they do not modify the qualitative picture, which is what is of interest in this part of the analysis.}: the increase in sensitivity does not change the relative behaviour between the QCD and same-charge bounds. It is noticeable that, for the 100\% $Wd$ case, where a comparison with the bound from~\cite{ATLAS:2024zlo} is possible, the ATLAS search still has a much higher reach, which justifies even more the exploration of  complementary same-charge production using a dedicated strategy. Also the bounds from \cite{Buckley:2020wzk} are stronger in the regions where they are applicable, however, such bounds do not consider the effects of large widths and interferences, which can sizeably affect the kinematics, as we will show in \cref{sec:features}.
It is finally worth noticing that the values of the couplings associated with the exclusion bounds (both opposite-charge $T\bar T$  and same-charge $TT$) are all within the perturbative range, as can be seen from the isolines in the figure.\\

\begin{figure}[h!]
\centering
\includegraphics[width=.49\textwidth]{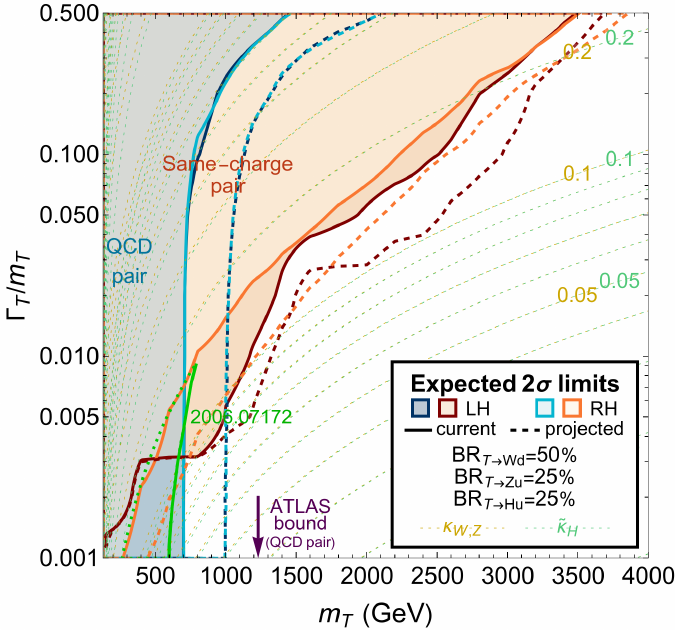}\hfill
\includegraphics[width=.49\textwidth]{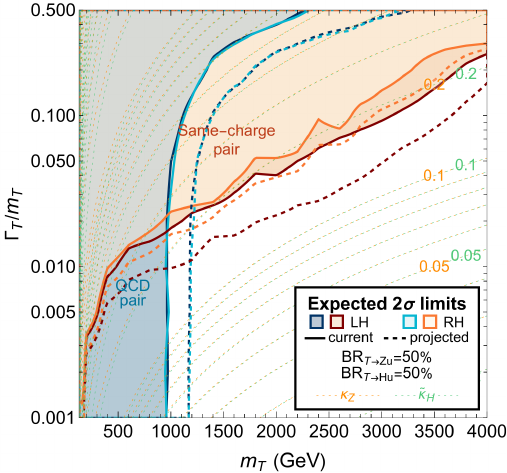}\hfill
\caption{\label{fig:singletvsdoublet}Same as \cref{fig:limitsmaximalBRs} but for the singlet-like ({left}) and doublet-like ({right}) scenarios. The comparison with the most recent ATLAS~\cite{ATLAS:2024zlo} bound and with \cite{Buckley:2020wzk} is possible only for the singlet-like case.}
\end{figure}

The three scenarios with 100\% BRs in one decay channel have the purpose to show the optimal limits in the three extreme cases, but offer limited scope to theoretical interpretation. Conversely, the singlet-like and doublet-like scenarios provide a more motivated description of BSM physics involving VLQs. In \cref{fig:singletvsdoublet} we show, for singlet-like and doublet-like scenarios, the complementary constraints from opposite-charge $T\bar T$  and same-charge $TT$ pair production.

Analogously to the 100\% BR scenarios, the same-charge $TT$  production dominates over opposite-charge $T\bar T$  production when the mass and width/mass ratio exceed specific values, namely, above $m_T\simeq650$ GeV and $\Gamma_T/m_T\simeq 0.005$ in the singlet-like case and $m_T\simeq750$ GeV and $\Gamma_T/m_T\simeq 0.02$ in the doublet-like case. In both cases, even if the QCD-driven bound is quite sensitive to the width in the large width limit and excludes much higher values of the $T$ mass, the constraint from the EW-driven process becomes competitive already when the width has relatively small values with respect to the mass, in the percent range. The exclusion from the same-charge pair production is very stringent when widths are large: masses around 3.5 TeV can be excluded for width/mass ratios of 50\% in the singlet-like case, while masses above 4 TeV can be excluded for width/mass ratios above 20\% in the doublet-like case. With the considered searches, the dependence of the bounds on the dominant chirality of the couplings is more visible in the low mass region for singlet-like case for the same-charge pair production process, even if for the left-handed case, fluctuations due to a harder phase-space integration play an important role. For the doublet-like case, the chirality dependence is slightly more pronounced in the high-mass, high-width region. \\

For the true singlet and true doublet case, let us first consider the exclusion determined by the APV observables. The weak charge of the nucleus can be written as~\cite{Deandrea:1997wk}:
\begin{equation}
Q_W={2 \cos \theta_w \over g}\left[(2Z+N)(g_{ZL}^u + g_{ZR}^u)+(Z+2N)(g_{ZL}^d + g_{ZR}^d)\right]\;,
\end{equation}
where $Z$ and $N$ are, respectively, the numbers of protons and neutrons in the nucleus, $g$ is the weak coupling constant, $\theta_w$ is the weak mixing angle and $g_{ZL,ZR}^{u,d}$ represent the LH and RH couplings of the SM $Z$ boson with up and down quarks, respectively.
The contribution of a $T$ VLQ is therefore~\cite{Okada:2012gy}:
\begin{equation}
\delta Q_W^T=2(2Z+N)\left((T_3^T-{1\over2})\sin^2\theta_L^T+T_3^T\sin^2\theta_R^T\right)\;,
\end{equation}
where $\theta_{L,R}^T$ are the LH and RH $T$ mixing angles and $T_3^T$ is its weak isospin.
The most accurate observations to date come from Cesium $^{133}$Cs ($Z=55, N=78$) and Thallium $^{204}$Tl ($Z=81, N=123$)~\cite{ParticleDataGroup:2024cfk}:
\begin{equation}
{\setlength{\arraycolsep}{1pt}
\begin{array}{lclcl}
Q_W^{\rm exp}(^{133}{\rm Cs})= - 72.41 \pm 0.42 &,~& Q_W^{\rm SM}(^{133}{\rm Cs})= - 73.26 \pm 0.01 &,~& \delta Q_W(^{133}{\rm Cs}) = 0.85 \pm 0.42\;,\\
Q_W^{\rm exp}(^{204}{\rm Tl})= - 116.4 \pm 3.6 &,~& Q_W^{\rm SM}(^{204}{\rm Tl})= - 116.93 \pm 0.01&,~& \delta Q_W(^{204}{\rm Tl}) = 0.53 \pm 3.6\;.
\end{array}
}
\end{equation}
While the Thallium results exhibit a good agreement, the Cesium results have a 2$\sigma$ discrepancy. 

In the case of Cesium APV, the contribution from $T$ could ease the tension, so that the 2$\sigma$ limit is given by $|\delta Q_W^T|>0.85 - 2\sigma$ or $|\delta Q_W^T|<0.85 + 2\sigma$, corresponding to $0.01<|\delta Q_W^T|<1.69$.
In the case of Thallium APV, we can place an upper bound on the mixing angle considering the 2$\sigma$ limit $|\delta Q_W^T|<0.53+7.2=7.73$.

These constraints translate in the following regions for the mixing angles of the two $T$ VLQ representations (LH for singlet, RH for doublet):\footnote{The contribution of the $B$ VLQ is absent at LO, due to the assumption of no mixing with the SM bottom quark its only interaction can be through a charged current with the SM top, originated by the $T-t$ mixing.}
\begin{eqnarray}
|\delta Q_W^T|=(2Z+N) \sin^2\theta_{L,R}^T \quad\Rightarrow\quad \left\{\begin{array}{ll}
0.0072 < |\sin\theta_{L,R}^T| < 0.095 & \text{allowed by }^{133}{\rm Cs~ APV}\\
|\sin\theta_{L,R}^T| < 0.16 & \text{allowed by }^{204}{\rm Tl~ APV}
\end{array}
\right.\;.
\end{eqnarray}
The Cesium upper limit is more stringent than the Thallium one. For what concerns the lower limit, the minimum value of the mixing angles in our simulation range are $\sin\theta_L^T=0.0098$ and $\sin\theta_R^T=0.014$, for the singlet and doublet, respectively, when $m_T=4$ TeV and $\Gamma_T/m_T=0.001$. Therefore, we will only consider the upper limit when comparing with LHC bounds, but it is interesting to notice that, if we want to explain the 2$\sigma$ discrepancy in $^{133}$Cs APV using exclusively a $T$ VLQ, a combination of large masses and small widths would not be sufficient.

\begin{figure}[h!]
\centering
\includegraphics[width=.49\textwidth]{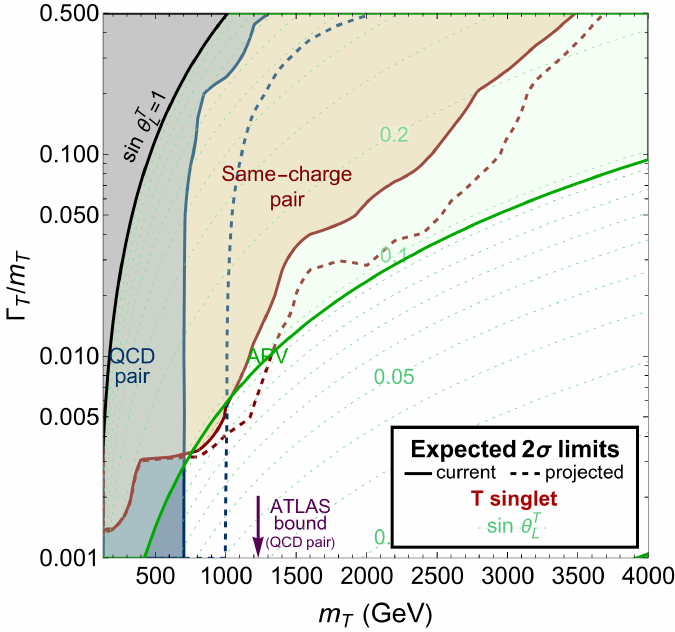}\hfill
\includegraphics[width=.49\textwidth]{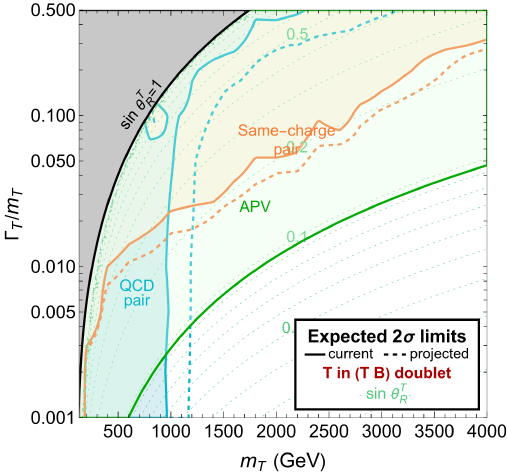}\hfill
\caption{\label{fig:truesingletvsdoublet}Same as \cref{fig:limitsmaximalBRs} but for the $T$ singlet ({left}) and $T$ in $(T~B)$ doublet ({right}) scenarios. The constraint from APV data in Cesium is also shown. The sine of the $T-t$ mixing angle is indicated by isolines.}
\end{figure}

In \cref{fig:truesingletvsdoublet} we show the cases of true singlet and true doublet, together with constraints from APV measurements. Notice that only the dominant chiralities corresponding to the interactions allowed by the VLQ representation under $SU(2)$ are shown for these scenarios. It is clear that, by  using the results from the recast of available searches, the bound coming from same-charge $TT$ production is largely subdominant with respect to the APV bound on the mixing angle, with the exception of the low mass range for the singlet case. However, the opposite-charge $T\bar T$  bound can be competitive, indeed, even dominant in the low width/mass ratio region for $T$ masses larger than $\sim$400 GeV for the singlet and $\sim$600 GeV for the doublet, respectively. In this area the mixing angle is too small to be constrained by APV and the $T$ masses are still light enough to be consistent with the 2$\sigma$ discrepancy in the APV measurement while still being within the LHC range. Considering the new ATLAS bound, however, and assuming an analogous improvement in the same-charge $TT$ production channel, the LHC constraints are likely to be competitive with the APV ones already with the full Run 2 data. Indeed, already the 300 fb$^{-1}$ projected bounds from the available searches show significant improvements: especially in the singlet case, the same-charge $TT$ production constraint will be competitive with the APV bound up to $m_T\simeq 2$ TeV already considering non-dedicated signal regions from the available recasts of LHC searches.

\subsection{Most sensitive signal regions and connection to current experimental bounds}
\label{sec:SRs}

\begin{figure}[h!]
\centering
\begin{minipage}{.49\textwidth}
\includegraphics[width=\textwidth]{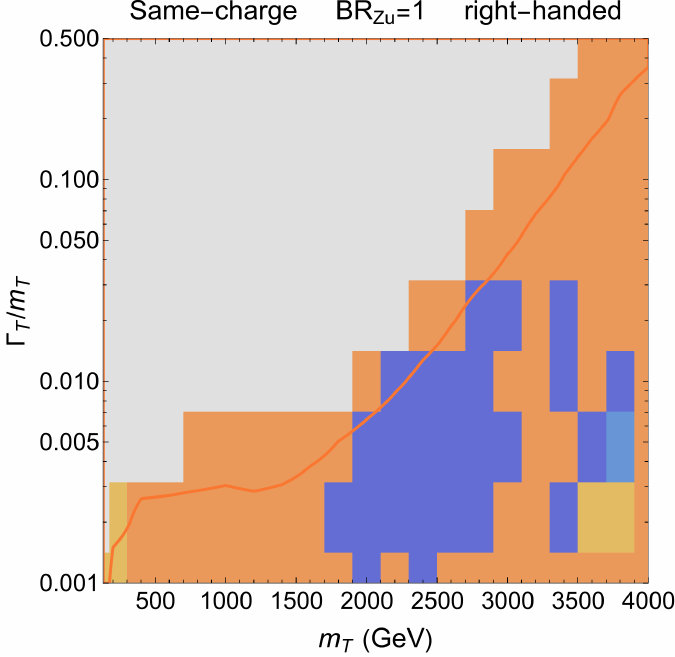}
\end{minipage}\hfill
\begin{minipage}{.49\textwidth}
\includegraphics[width=\textwidth]{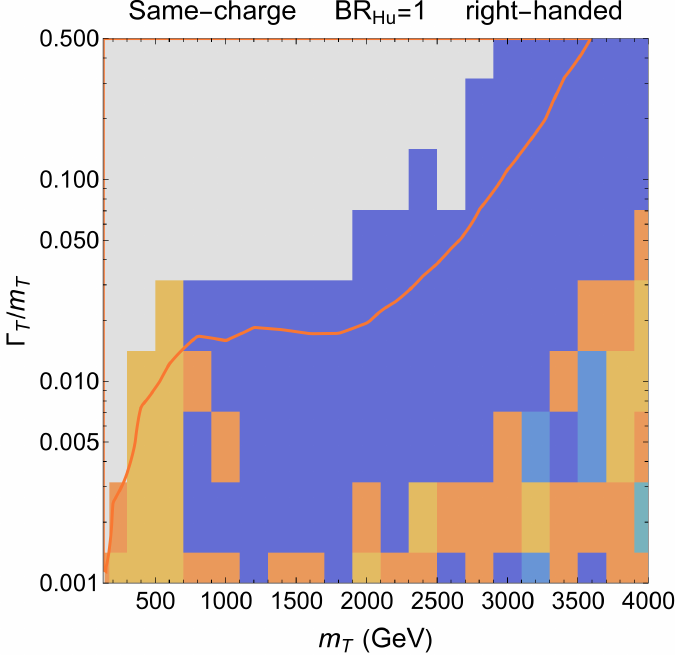}
\end{minipage}

\begin{minipage}{.49\textwidth}
\includegraphics[width=\textwidth]{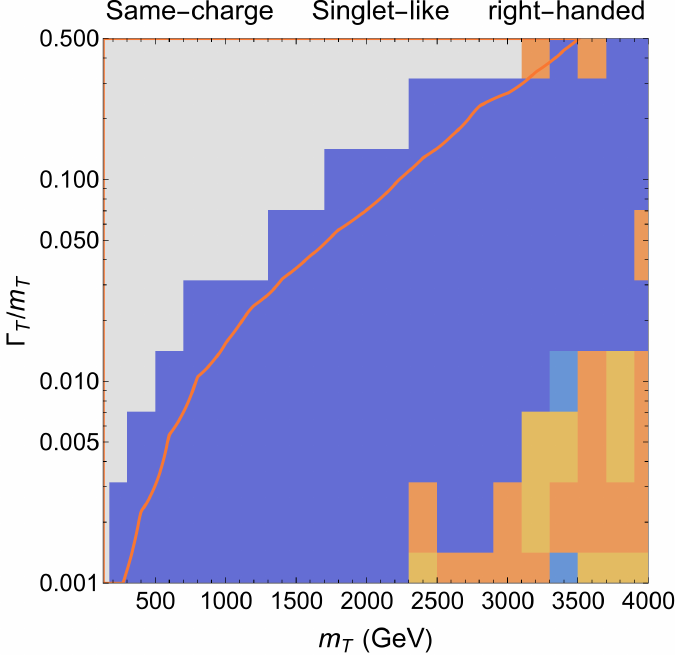}\hfill
\end{minipage}\hfill
\begin{minipage}{.49\textwidth}
\hspace*{35pt}\includegraphics[width=.82\textwidth]{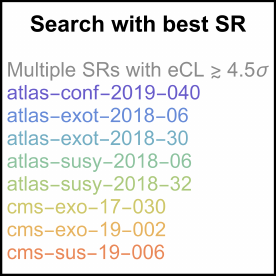}\\
\end{minipage}
\caption{\label{fig:bestSRs} Searches with the most constraining signal regions for different benchmarks: only $Zu$ coupling {(top left)}, only $Hu$ coupling {(top right)} and singlet-like scenario {(bottom left)}, all with RH couplings. The legend reports all the searches used in the recast analysis.}
\end{figure}

The bounds presented in the previous section are associated to the different sensitivity of the various searches we used to the different objects in the final states and their kinematical properties, depending on the mass, width, coupling chiralities and BRs of the $T$. 
In this section we will explore in more detail which signal regions (SRs) exhibit the highest sensitivity for the same-charge process and for three representative scenarios, namely when the $T$ interacts only with $Zu$, only with $Hu$ and when it is singlet-like. The first two scenarios will show which SRs are more sensitive to the presence of $Z$ or $H$ in the final state, while the third one is largely sensitive to final states with $W$ bosons, as it will be clearer in the next section. Our results are shown in \cref{fig:bestSRs}, where the most sensitive searches are represented by colour codes. In this section we only consider scenarios with dominantly RH couplings. Including the results from LH couplings would not significantly change the main conclusions. 

Due to the limited numerical accuracy of the CLs algorithm, exclusion confidence level can be computed up to 4.5$\sigma$, and therefore if more signal regions provide a higher exclusion level, we have not identified a best one; in any case, we are interested in the region which can still be explored, which is the area around the 2$\sigma$ exclusion.   Barring statistical fluctuations, rather distinct patterns emerge:
\begin{itemize}

\item For the scenario with exclusive $Zu$ interaction, up to $m_T\sim$1.5--1.8 TeV the most sensitive search is \cite{CMS:2019zmd}, which targets final states with multiple jets, no leptons, and large missing transverse energy (MET). The best signal regions are bin 35 (see Table 4 in the CMS publication) for $m_T$ up to 1 TeV and the aggregate bin 2 (CMS Table 9) for larger $m_T$. Both SRs are characterised by the presence of at least 4 jets; bin 35 requires at most 5 jets, $H_T$ (defined as the scalar sum of transverse momenta of all jets) between 600 GeV and 1200 GeV and MET between 350 GeV and 600 GeV, while the aggregate bin 2 requires $H_T>1700$ GeV and MET$>$850 GeV. 

For larger mass values the most sensitive search can also be \cite{ATLAS:2020syg}, which is also a search targeting final states with mutiple jets, no leptons and large MET. The SR with highest sensitivity is SR-4j-3400 (see Table 8 in the ATLAS paper), characterised by the presence of at least 4 jets and a high 'effective mass', $m_{\rm eff}$, defined as the scalar sum of the transverse momenta of all jets and missing transverse momentum. Clearly this SR  is sensitive to new physics with large mass scales and global variables such as $m_{\rm eff}$ are indeed very useful to test processes involving heavy VLQs (see also \cite{Banerjee:2023upj}).

\item When $T$ interacts exclusively with the Higgs boson, the ATLAS search \cite{ATLAS:2020syg} still provides the highest sensitivity for masses above 1 TeV exploiting SRs with at least 4 jets and increasing cuts on the $m_{\rm eff}$ variable. However, for low $T$ masses, the search with the best sensitivity is mostly \cite{CMS:2019lwf}, which targets final states with multiple leptons. The SRs which maximally constrain the signal require 3 or 4 leptons, 1 b-jet and $S_T$, defined as the scalar sum of transverse momenta of jets, leptons and MET, larger than 800 GeV (see Table 2 in the CMS publication). This corresponds to one Higgs decaying to bottom quarks and the other leptonically.

\item In the singlet-like (and also true singlet) case, largely dominated by the $WdWd$ final state, the most sensitive search is always \cite{ATLAS:2020syg} through SRs with at least 2 jets and different cuts on the $m_{\rm eff}$ variable, due to the small multiplicity of objects in the final state.

\item In the doublet-like case, the mixture between $Zu$ and $Hu$ decays does not allow identifying regions where one search is more sensitive. The best SRs fluctuate between the three searches mentioned above in the whole scanned parameter space.

\end{itemize}

A discussion is now in order to relate the constraints from our recast to the bound from the recent ATLAS search~\cite{ATLAS:2024zlo}. This search specifically targets signals coming from $T$ decaying to $Wd$. It preselects events by identifying large- and small-$R$ jets, one lepton and MET, in order to reconstruct 2 $W$ bosons, one decaying hadronically and one leptonically. On top of that, it uses a strong cut on the $S_T$ variable defined above by requiring $S_T>2$ TeV. Two SRs are then defined, using complementary cuts on the azimuthal angle between jets and MET. This design clearly allows being sizeably more sensitive to the $WdWd$ final state, as also shown in our plots. It is reasonable to expect that the same-charge process would also be effectively tested by this search.

To conclude this part of the analysis, it is interesting to notice that the use of global variables such as $S_T$ or $m_{\rm eff}$ is one of the key elements for probing signals coming from $T$ decaying to light quarks: SRs which are generic enough, such as those based on global cuts and selections over the multiplicity of jets or leptons, already have the power to strongly constrain the same-charge process, which features a rich variety of final states depending on the assumptions about $T$ interactions. Still, optimisations of SRs to be sensitive to other $T$ decay channels and the peculiar features of the same-charge $TT$ pair production process will allow probing a much larger parameter space in the $\{m_T, \Gamma_T/m_T\}$ plane. This aspect will be discussed in the following section.

\subsection{Relevant features of the same-charge $TT$ production signal}
\label{sec:features}

As seen in the previous discussion, the same-charge $TT$ production signal differs from the opposite-charge $T\bar T$ QCD production in a few main aspects. 

\subsubsection{Cross-sections and signs of couplings} 

While the scaling of cross-sections for each subprocess is known analytically and shown in \cref{eq:sigmahats}, the scaling of the total cross-section crucially depends on the relative signs of the interference terms. The overall scaling of the cross-sections is therefore non-trivial and can change depending on the size and signs of the couplings for specific benchmark choices. Therefore, a numerical approach is needed to compare same-charge and QCD pair production processes for different masses and widths. Let consider as an example the singlet-like case with RH coupling and with $m_T=1$ TeV and $\Gamma_T/m_T=0.01$ and 0.1 respectively, {\it i.e.}, one point which is not excluded by either process (within our recast) but is in the reach of both at the end of Run 3, and one point which is excluded by same-charge production but not by QCD production. The couplings which produce the correct BRs with $2:1:1$ pattern are $k^W=k^Z\simeq0.123$ (0.39) and $k^H\simeq0.509$ (1.61) for $\Gamma_T/m_T=0.01$ (0.1). With these values, the cross-sections are:
\begin{eqnarray}
\begin{array}{l}
\sigma_{TT}(1~{\rm TeV}, 0.01)\simeq 57~{\rm fb} \\
\sigma_{T\bar T}(1~{\rm TeV}, 0.01)\simeq 36~{\rm fb}
\end{array}\quad
\begin{array}{l}
\sigma_{TT}(1~{\rm TeV}, 0.1)\simeq 979~{\rm fb} \\
\sigma_{T\bar T}(1~{\rm TeV}, 0.1)\simeq 99~{\rm fb}
\end{array}\;,
\end{eqnarray}
where the breakdown of the contributions leading to these values is shown in \cref{app:breakdownxs}. The cross-section for same-charge pair production is larger than the one for QCD pair production of a factor 1.6 when $\Gamma_T/m_T=0.01$ and 10 when $\Gamma_T/m_T=0.1$, showing that for any given mass, as couplings (and thus widths) increase, same-charge production becomes more and more dominant with respect to opposite-charge QCD production.\\

The previous results, including the bounds obtained in \cref{sec:results}, rely on the assumption that the signs of the $\{\kappa^W,\kappa^Z,\kappa^H\}$ couplings of \cref{eq:LagTSM} are the same, either all positive or all negative. Part of the interference terms, however, depend on odd powers of the couplings, and therefore, if the signs are discordant, the same topology can increase or decrease the total cross-section. In \cref{tab:xssigns} we report the values of the cross-sections for the same benchmark discussed above, but for varying coupling chiralities and relative signs.\footnote{Complementary sign configurations such as $++-$ and $--+$ are equivalent: since we are treating a $2\to4$ process, the amplitude of each contribution is proportional to the product of 4 couplings, but since the Lagrangian contains only 3 free couplings, if one of them enters the amplitude with odd power, another coupling must enter with odd power too, so that if both change sign the cross-section does not change.} The values can be extremely different, especially when the dominant chirality is LH, for which interference terms with the SM background have a much larger relevance. These differences can clearly strongly affect the constraints one can pose on scenarios featuring VLQs with same masses and total widths and can affect the design of analysis targeting these particles. A detailed analysis of the dependence of current bounds and of the kinematical properties of the final state on the sign of the couplings is postponed to a forthcoming analysis.
\begin{table}[h!]
\centering
\begin{tabular}{ccc|cccc}
\toprule
\multirow{2}{*}{process} & \multirow{2}{*}{$\Gamma_T/m_T$} & \multirow{2}{*}{chirality} & \multicolumn{4}{c}{cross-sections (fb) depending on $\{\kappa^W,\kappa^Z,\kappa^H\}$ signs}\\[-5pt]
&&& $+++/---$ & $+--/-++$ & $++-/--+$ & $+-+/-+-$ \\
\midrule
\multirow{4}{*}{\vspace*{5pt}{same-charge}} & \multirow{2}{*}{0.01} & LH & 83.544 & 48.875 & 41.053 & 75.690 \\[-5pt]
&& RH & 57.440 & 57.440 & 60.887 & 60.887 \\
\cmidrule{2-7}
& \multirow{2}{*}{0.1}  & LH & 520.301 & 1327.78 & 131.122 & 1563.5  \\[-5pt]
&& RH & 979.495 & 979.495 & 1012.77 & 1012.77 \\
\midrule
\multirow{4}{*}{\vspace*{5pt}{QCD}} & \multirow{2}{*}{0.01} & LH & 38.935 & 37.120 & 33.827 & 36.005 \\[-5pt]
&& RH & 35.729 & 35.729 & 37.085 & 37.085 \\
\cmidrule{2-7}
& \multirow{2}{*}{0.1}  & LH & 130.379 & 112.203 & 79.687 & 101.052 \\[-5pt]
&& RH & 98.512 & 98.512 & 111.99 & 111.99 \\
\bottomrule
\end{tabular}
\caption{\label{tab:xssigns} Cross-sections for a singlet-like $T$ (${\rm BR}_{T\to Wd}$:${\rm BR}_{T\to Zu}$:${\rm BR}_{T\to Hu}$=2:1:1) with mass $m_T=1$ TeV and two values of the total width/mass ratio, namely 0.01 and 0.1 depending on the coupling chiralities and on the relative signs of the $\{\kappa^W,\kappa^Z,\kappa^H\}$ couplings in \cref{eq:LagTSM}.}
\end{table}

\subsubsection{Final states} 
If the $T$ decays to $Zu$ or $Hu$, the observable objects in the final state are clearly not different between same-charge and QCD pair production, as they would both involve the presence of $Z$ or $H$ decay products and two jets. For these decay channels only the cross-section and the kinematics of the final state can be exploited to design dedicated searches for same-charge production. However, if the $T$ decays through a charged-current to $Wd$, it is in principle possible to exploit the presence of same-sign leptons in the final state. The high cross-sections corresponding to large values of the width allow to not deplete the signal excessively when selecting this final state. 

Let's consider again the singlet-like scenario with RH couplings and $m_T=1$ TeV: from the breakdown of contributions reported in \cref{app:breakdownxs}, it is possible to see that the $WdWd$ final state contributes to 97\% of the cross-section when the width/mass ratio is 0.01 and 77\% when the ratio is 0.1. These proportions are very large, even if the BR into $Wd$ is 50\%. To explain these values, notice that the same-charge production cross-section always depends on the couplings to $Z$ and $H$, and therefore interference contributions can play a relevant role, especially when the width is small (see also the discussion at the end of~\cite{Carvalho:2018jkq}): indeed, in the case under consideration, the contribution of $\sigma^{ZB_{\rm int}}$ to the cross-section in the $WdWd$ final state is largely dominant when the width/mass ratio is 0.01, while $\sigma^Z$ is comparable when the ratio is 0.1. Therefore, the proportion of events going in different final states has a non-trivial dependence on the $T$ mass and total width for a given choice of BRs. This is in contrast with the QCD pair production case, where only the couplings of $T$ with the final states matter, and therefore the proportion of events in the $WdWd$ final state is closer to 25\% when the width is small, in accordance with the singlet-like BR relations. 

For a point with $m_T=1.5$ TeV, width/mass ratio of 2\% and RH couplings, still not excluded even by the most recent ATLAS search~\cite{ATLAS:2024zlo}, our numerical results show that the total cross-section is $\sim 24$ fb with a $WdWd$ proportion of around 95\%. Now, considering that the $W$ BRs into electrons and muons (including the contribution of $\tau$ decays into $\ell \nu \bar \nu$) is $\sim$25\%, we can compute the effective cross-sections as:
\begin{eqnarray}
TT\to 2\ell^\pm+2j+MET: \quad 
\left\{
{
\setlength{\arraycolsep}{2pt}
\begin{array}{l}
\sigma^{\rm eff}(m_T=1~{\rm TeV}, \Gamma_T/m_T=0.01) \simeq 3.4~{\rm fb}\\
\sigma^{\rm eff}(m_T=1.5~{\rm TeV}, \Gamma_T/m_T=0.02) \simeq 1.4~{\rm fb}
\end{array}
}
\right.\;,
\end{eqnarray}
which are large enough to be effectively tested at Run 3, considering the very small SM background for this final state~\cite{Cui:2022hjg}.

\subsubsection{Kinematics} 

In this section, we will show the main kinematical differences between same-charge and QCD pair production processes, and also include the dependence on the chirality of the coupling. To be concrete, let's consider the same parameters used above, {\it i.e.}, a singlet-like $T$ with $m_T=1.5$ TeV and $\Gamma_T/m_T=0.02$ and analyse the distributions of events for the final Run 3 luminosity of 300 fb$^{-1}$. The distributions are shown for reconstructed objects after hadronisation, parton showering and decays of SM bosons both before any selection. \\

\begin{figure}[h!]
\centering
\begin{minipage}{.49\textwidth}
\includegraphics[width=\textwidth]{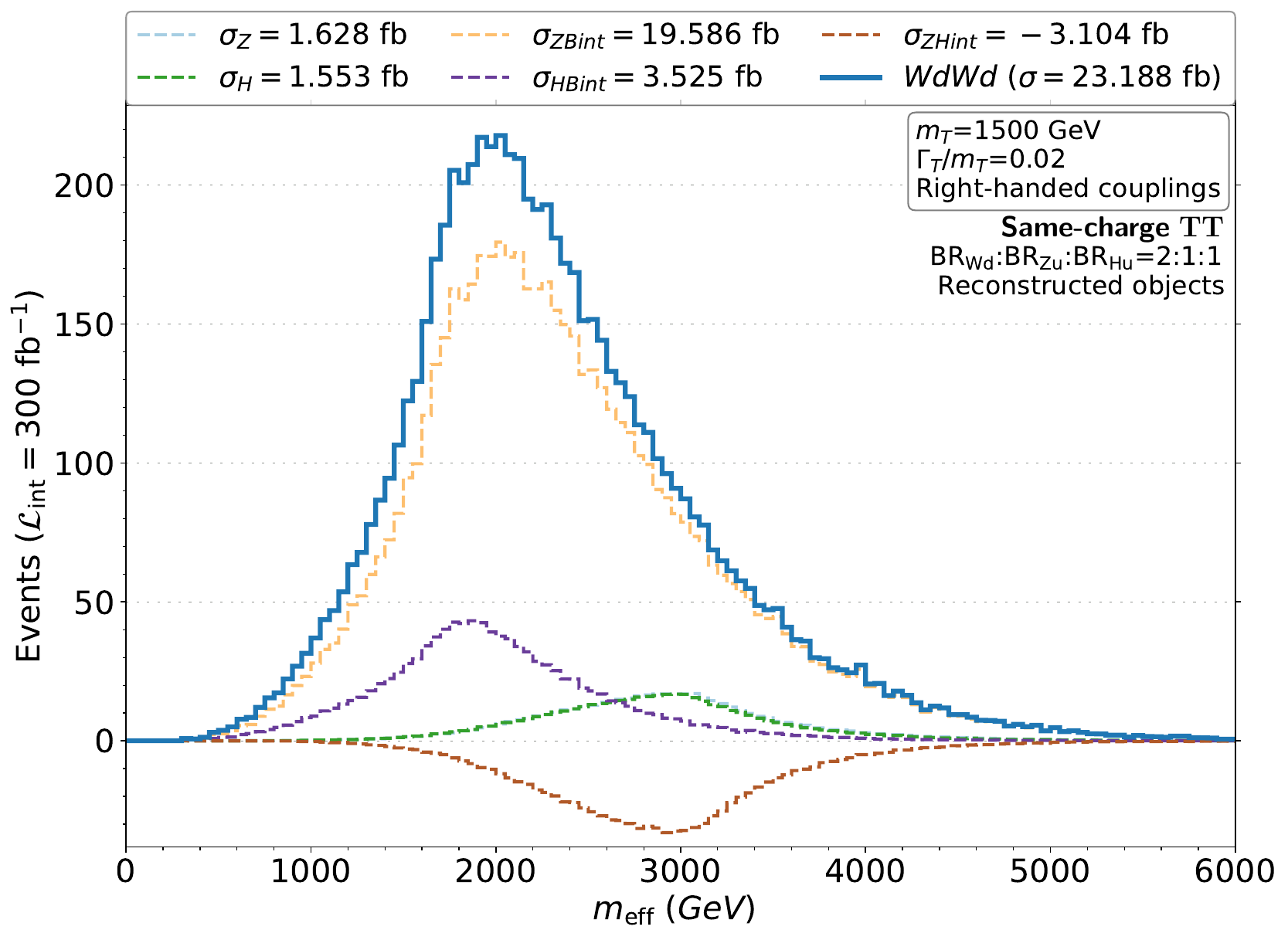}
\end{minipage}\hfill
\begin{minipage}{.49\textwidth}
\includegraphics[width=\textwidth]{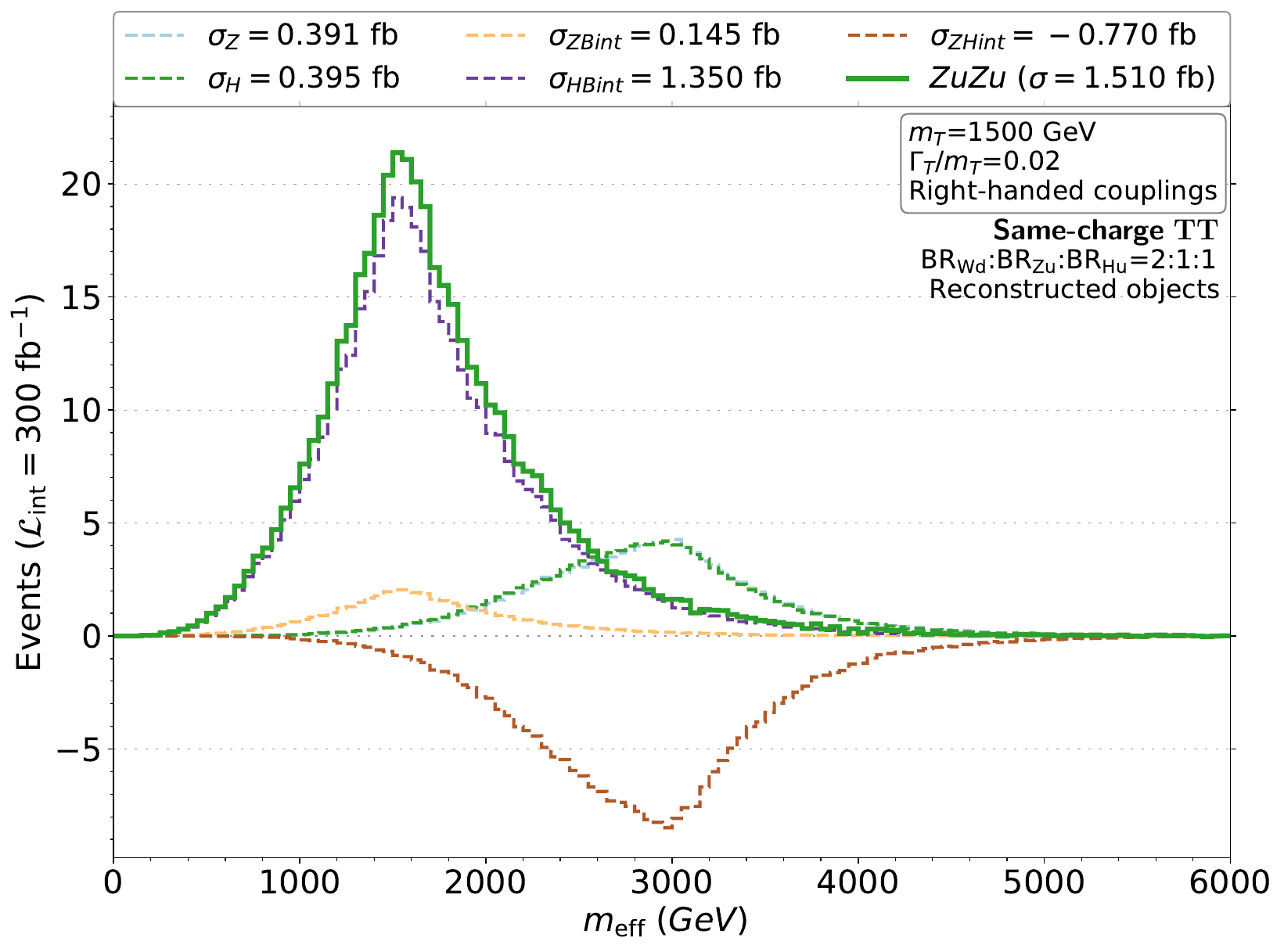}
\end{minipage}
\caption{\label{fig:procdeconstruction} Contributions of $Z$- and $H$-mediated subprocesses to the distributions of parton-level events corresponding to a luminosity of 300 fb$^{-1}$ in the $m_{\rm eff}$ observable, for the $WdWd$ ({left}) and $ZuZu$ ({right}) final states in the same-charge $TT$ production process of a singlet-like $T$ with $m_T=1.5$ TeV, $\Gamma_T/m_T=0.02$ and RH couplings. Negative values are unphysical {\it per se}, but contribute as a {\it depletion} of events once the irreducible background is added to the signal to produce a positive (or null for maximal negative interference) quantity.}
\end{figure}

The various elements of \cref{eq:sigmahats} all contribute (with different weights) to the kinematics of the objects in the final state. To make their role explicit, let's focus on two representative final states for the same-charge production, $WdWd$ and $ZuZu$. For both of them, we show in \cref{fig:procdeconstruction} the contributions of the five terms of  \cref{eq:sigmahats} to the $m_{\rm eff}$ distribution at object-reconstruction level ({\it i.e.}, after decay of SM bosons, hadronisation and parton showering but without detector effects), where this quantity is defined as the scalar sum of the transverse momenta of final state visible objects and missing transverse momentum. 

As expected, the contributions associated with $\sigma_Z$ and $\sigma_H$, and their interference, peak around twice the mass of the $T$, while interference terms with the irreducible SM background have a maximum at lower energy. It is also noticeable how the peak of the interference terms for $ZuZu$ are around 1.5 TeV, signalling a large relevance of topologies with one resonant $T$ (see left diagram in \cref{fig:LWtopologies}). In both cases the cancellation of opposite contributions from $\sigma_Z+\sigma_H$ and $\sigma_{ZH}^{\rm int}$ is almost exact also at differential level, and the shape of the distributions is entirely determined by the interferences with the SM. Even with a relatively small width, therefore, the NWA approach, which only considers the $\sigma_Z+\sigma_H$ contributions, would not describe correctly both the total cross-section and the kinematics of these final states. \\

\begin{figure}[h!]
\centering
\begin{minipage}{.49\textwidth}
\includegraphics[width=\textwidth]{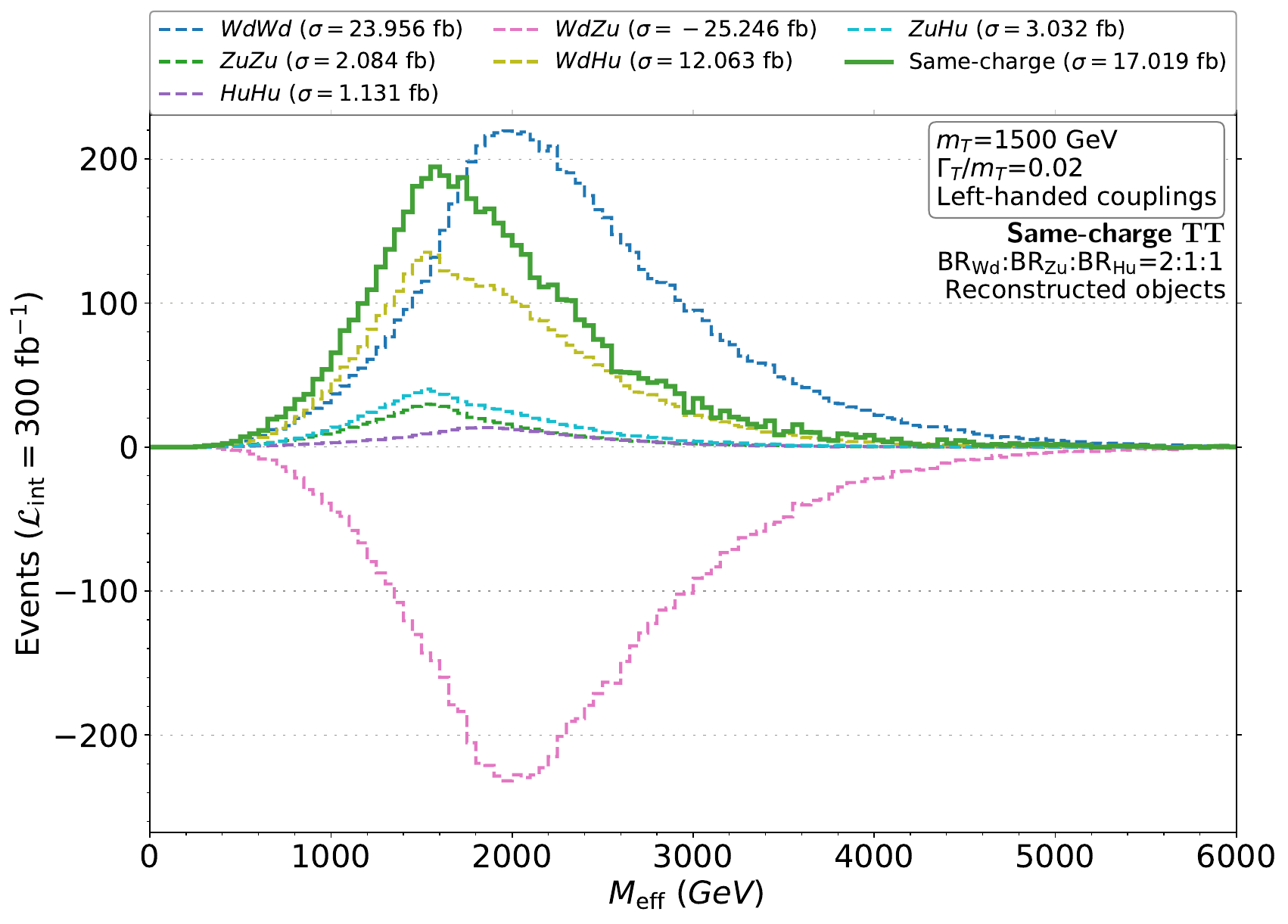}
\end{minipage}\hfill
\begin{minipage}{.49\textwidth}
\includegraphics[width=\textwidth]{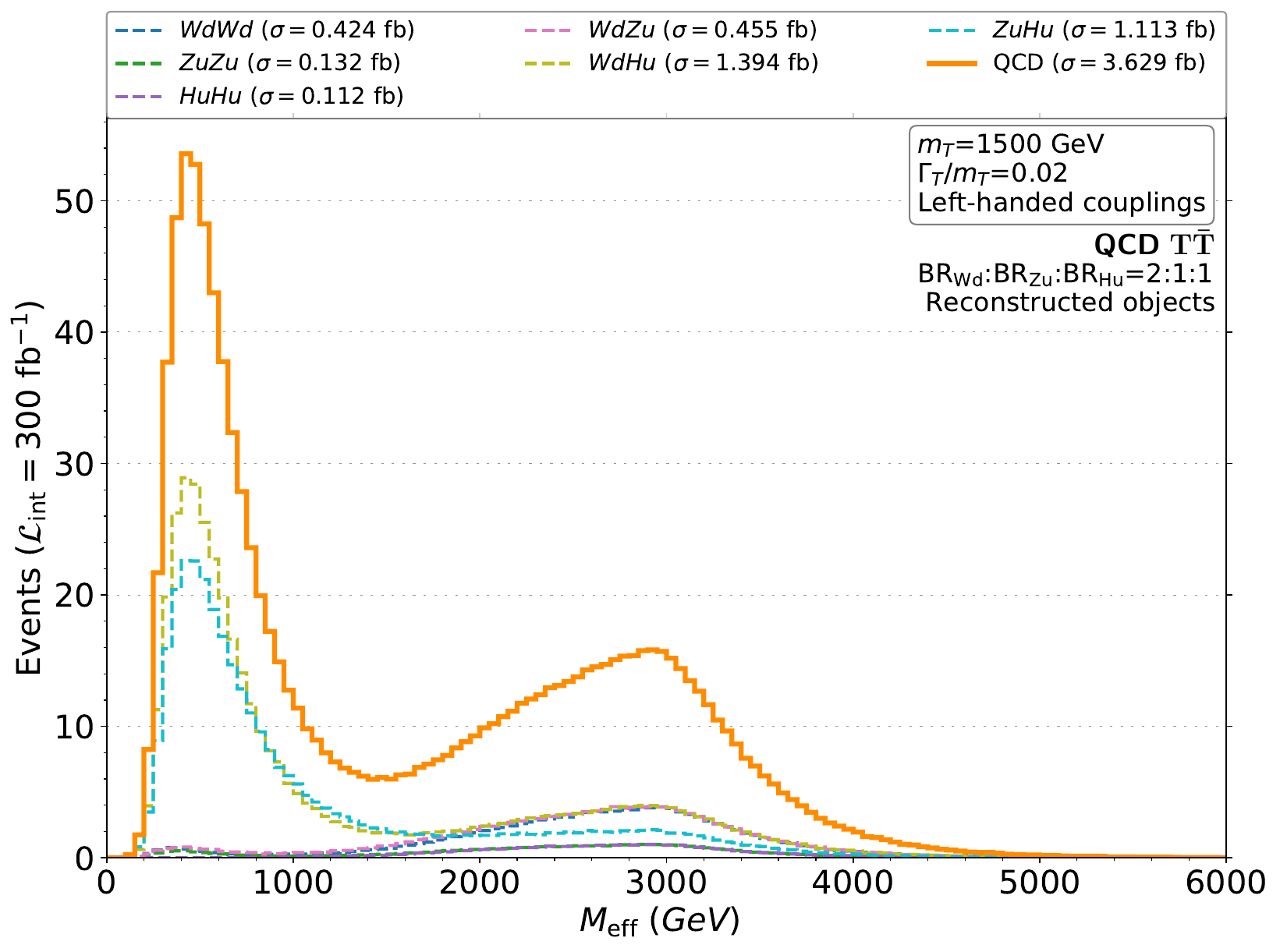}
\end{minipage}\\
\begin{minipage}{.49\textwidth}
\includegraphics[width=\textwidth]{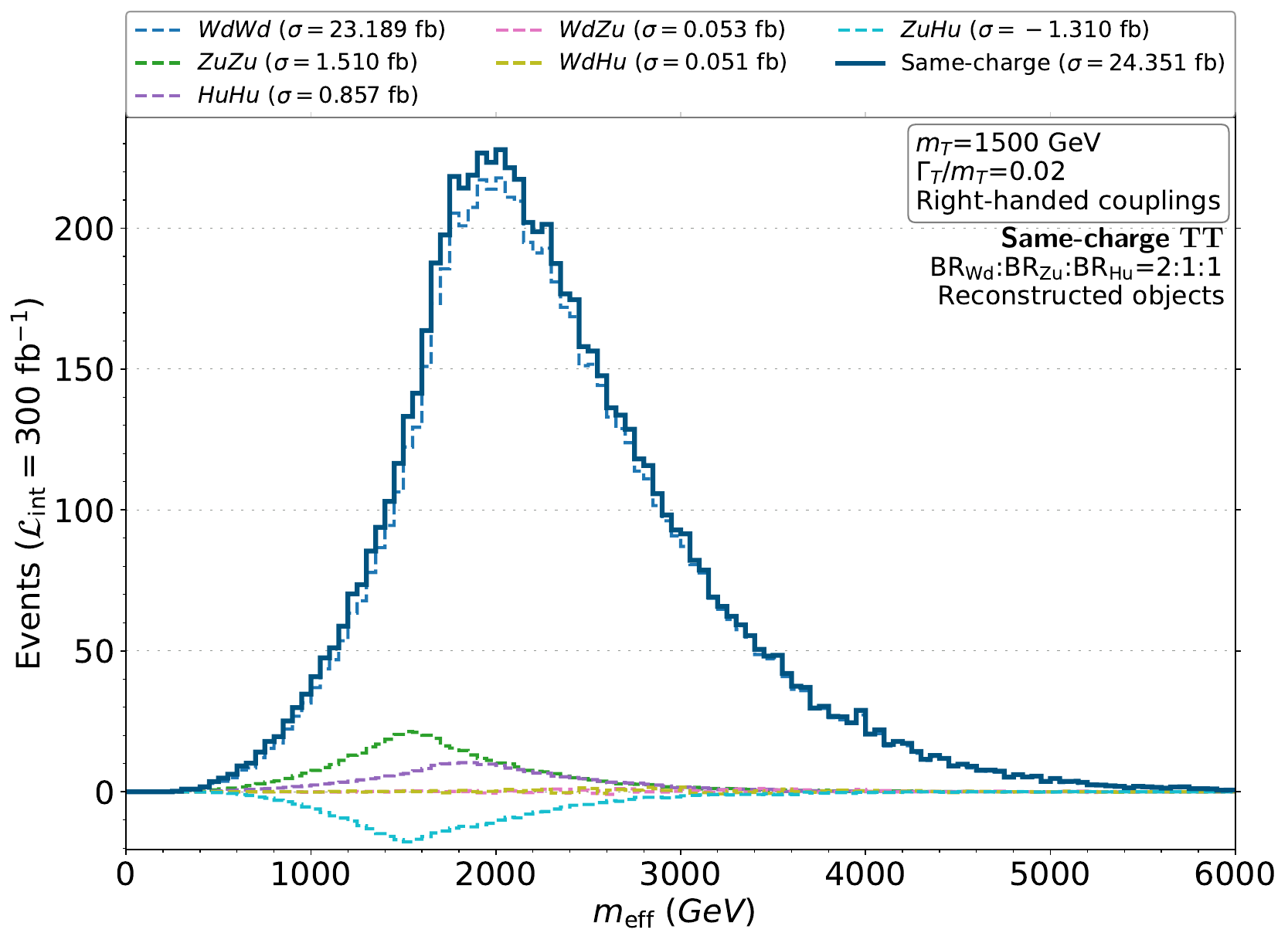}
\end{minipage}\hfill
\begin{minipage}{.49\textwidth}
\includegraphics[width=\textwidth]{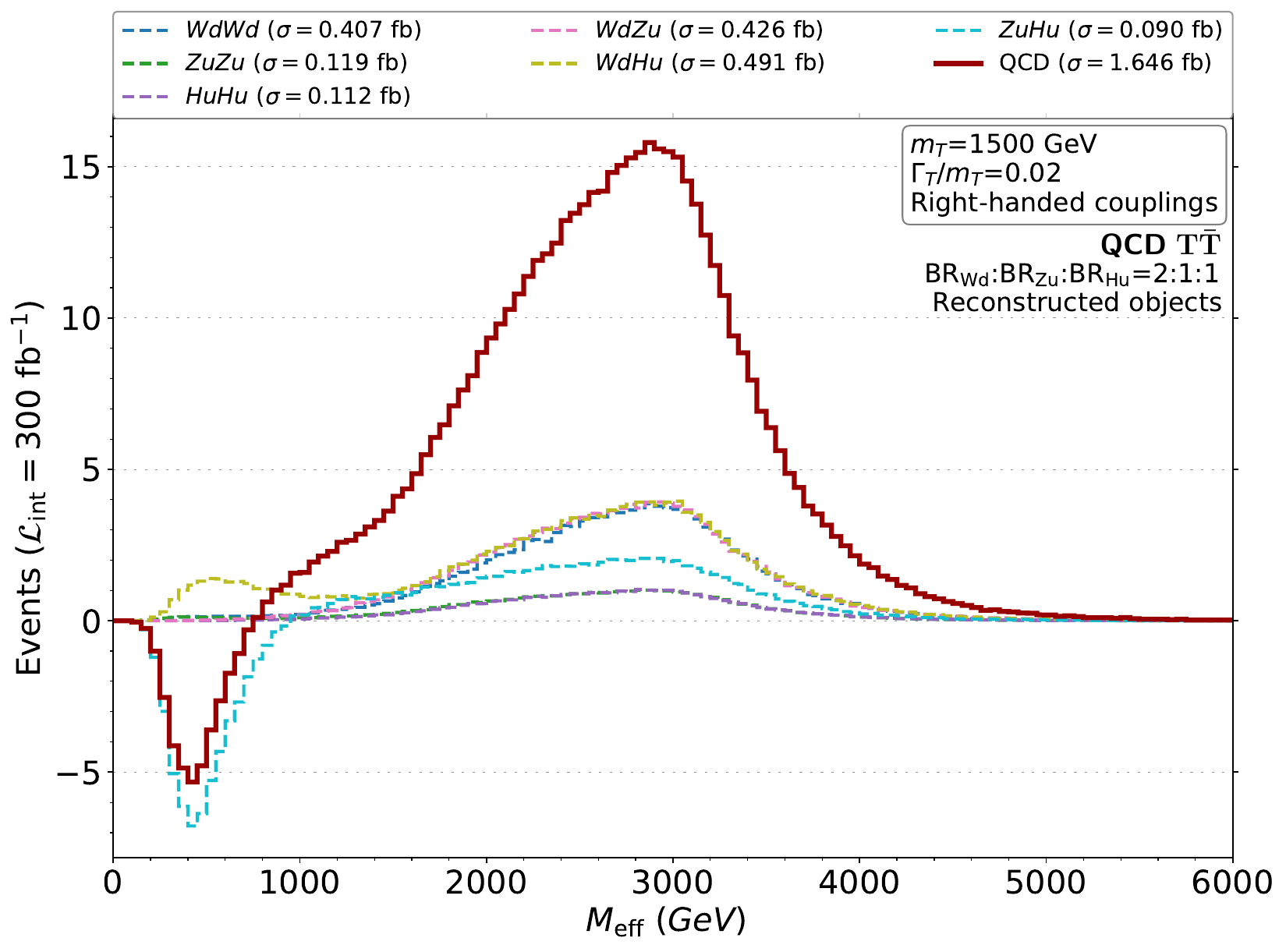}
\end{minipage}
\caption{\label{fig:finalstatedeconstruction} Contributions of different final states to the $m_{\rm eff}$ distribution for the same-charge ({left}) and QCD ({right}) production process of a singlet-like $T$ with $m_T=1.5$ TeV, $\Gamma_T/m_T=0.02$ with LH ({top}) and RH ({bottom}) couplings. For reference, in the same-charge with RH coupling case the contributions of $WdWd$ and $ZuZu$ correspond to the solid curves in \cref{fig:procdeconstruction}.}
\end{figure}

We can now repeat the same procedure for all the 6 possible final states corresponding to the interactions of the $T$ and for both same-charge and QCD production processes and for LH and RH couplings, and add all contributions to obtain the final differential distributions. We show the outcome of this operation in \cref{fig:finalstatedeconstruction}.

Clearly, the role of interference contributions is crucial in all cases. In the LH coupling scenario, for the same-charge production the positive contribution of the $WdWd$ process is almost entirely cancelled by the negative contribution of the $WdZu$ process at differential level, leaving as a net result the sum of all other contributions. For QCD production the interferences in the low energy region from the $WdHu$ and $ZuHu$ processes with the SM background dominate the shape of the differential distribution, while in the $m_{\rm eff}=2m_T$ region all contributions contribute analogously.

In the RH scenario, for same-charge production the dominant contribution is by far from the $WdWd$ final state, which contributes almost entirely to shape the differential distribution. In the QCD production case, in the low $m_{\rm eff}$ region the $WdHu$ contribution is much smaller and the $ZuHu$ contribution is negative, leading to an overall small {\it depletion} of events in the signal, while in the $m_{\rm eff}=2m_T$ region the shape is similar to the LH case.\\ 

\begin{figure}[h!]
\centering
\begin{minipage}{.49\textwidth}
\includegraphics[width=\textwidth]{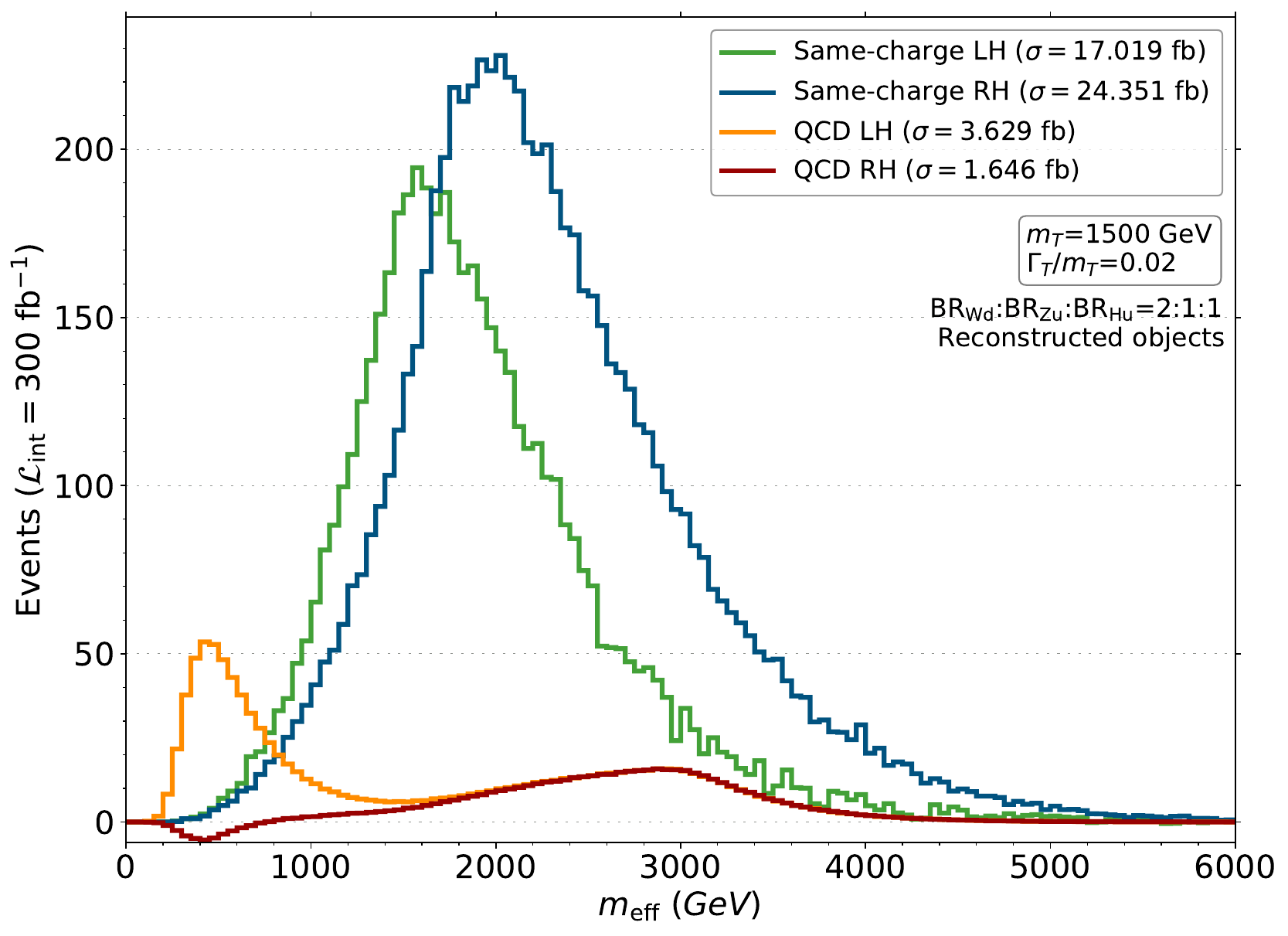}
\end{minipage}\hfill
\begin{minipage}{.49\textwidth}
\includegraphics[width=\textwidth]{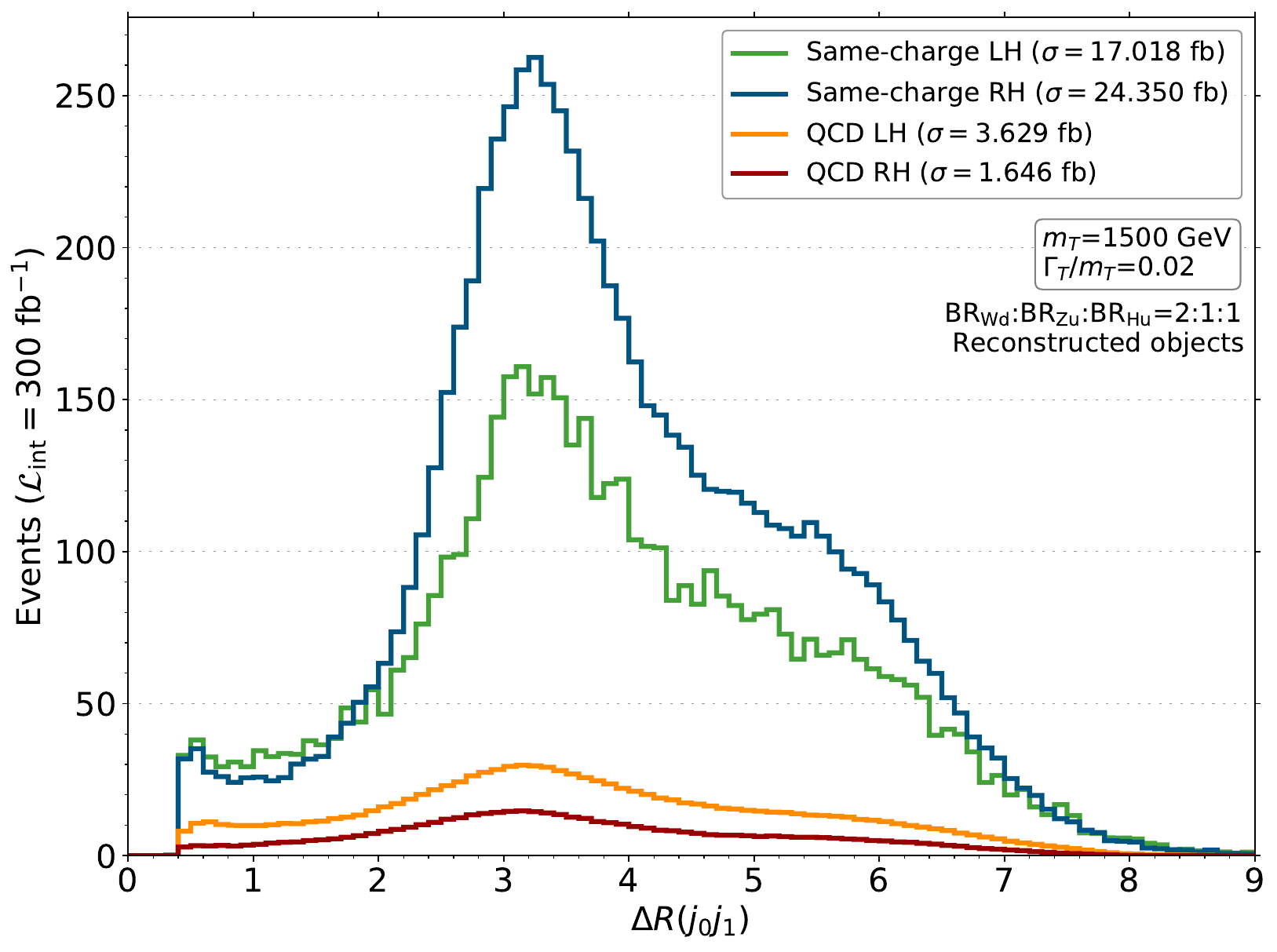}
\end{minipage}\\
\begin{minipage}{.49\textwidth}
\includegraphics[width=\textwidth]{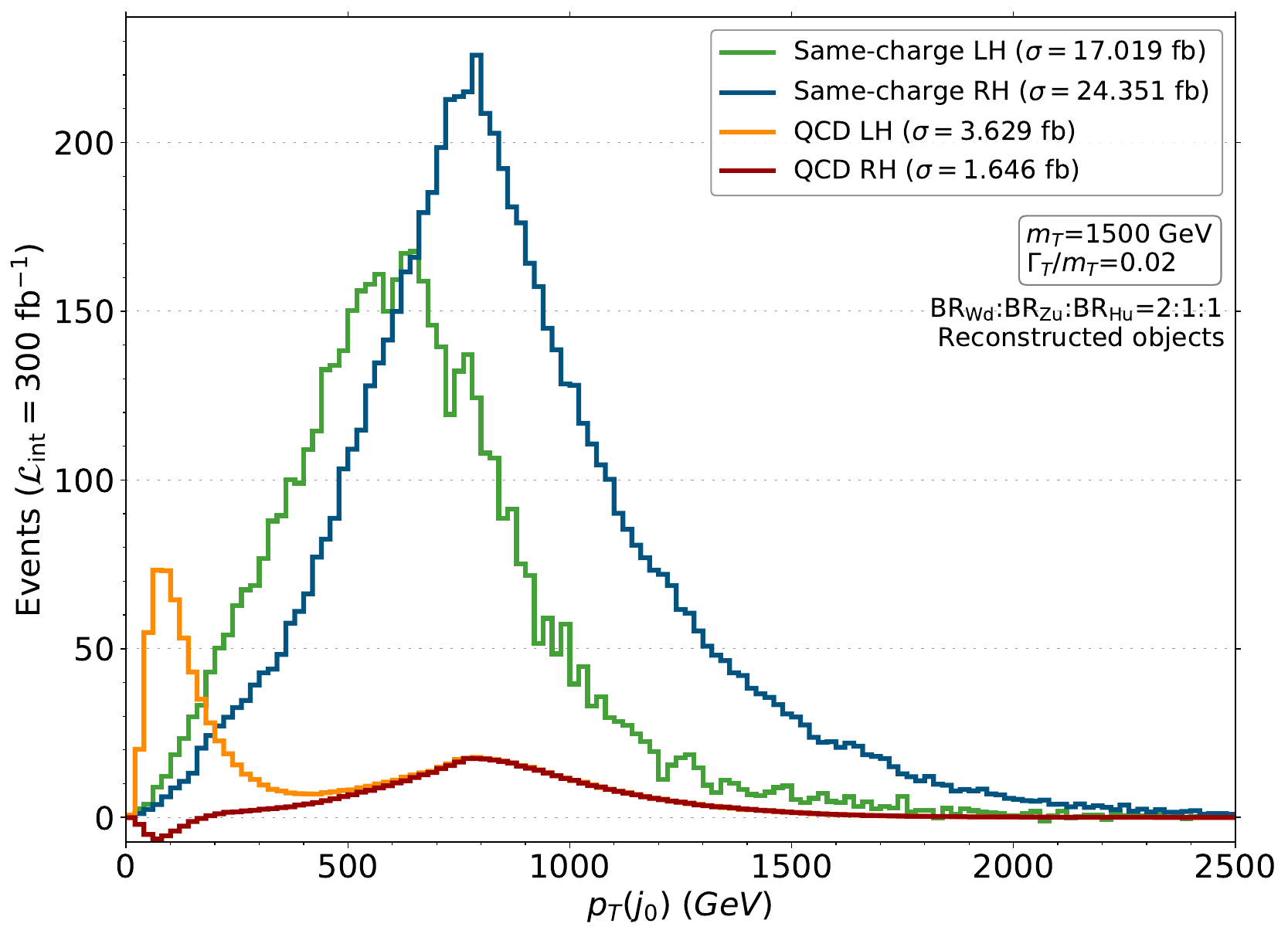}
\end{minipage}\hfill
\begin{minipage}{.49\textwidth}
\includegraphics[width=\textwidth]{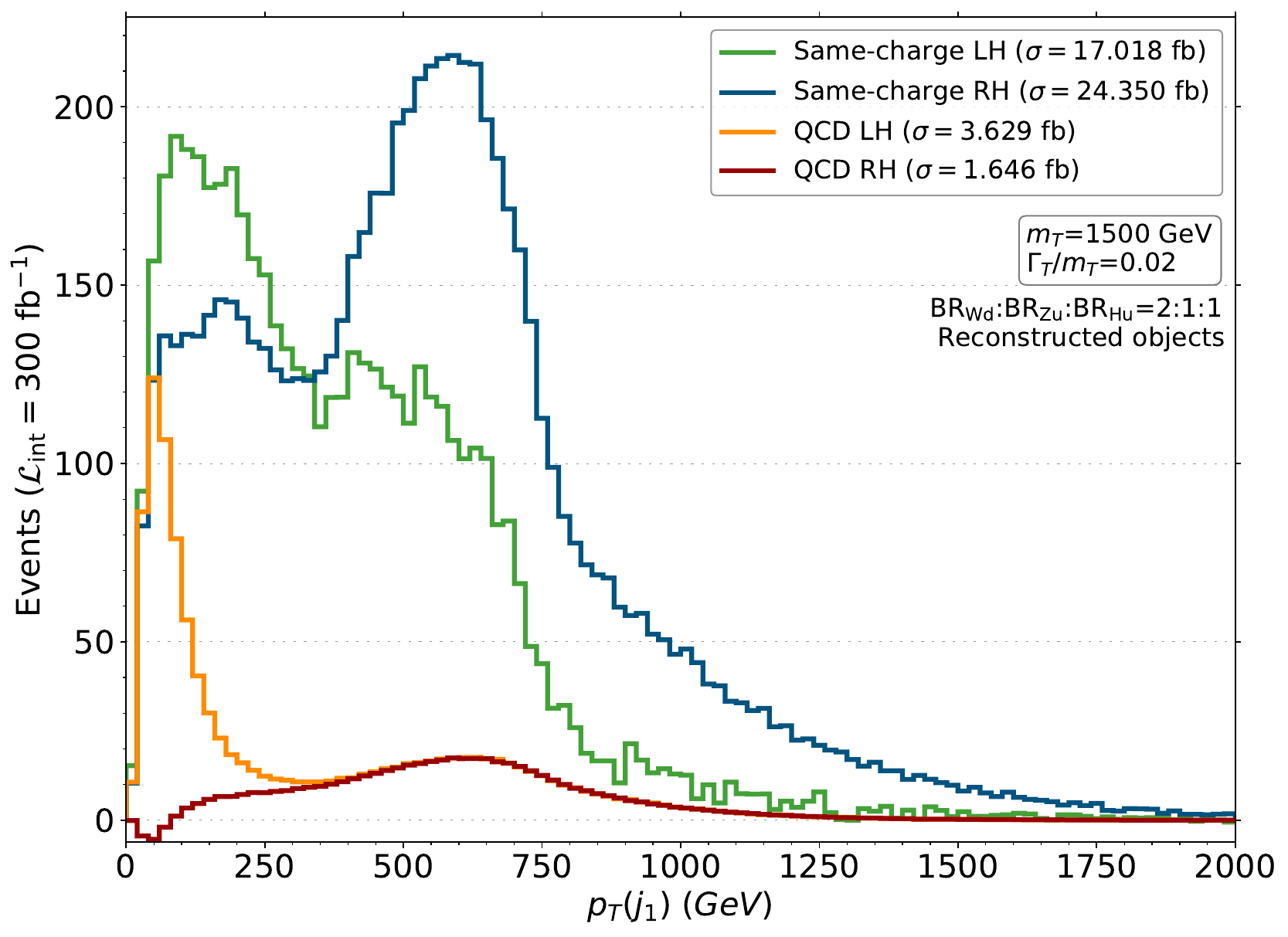}
\end{minipage}\\
\begin{minipage}{.49\textwidth}
\includegraphics[width=\textwidth]{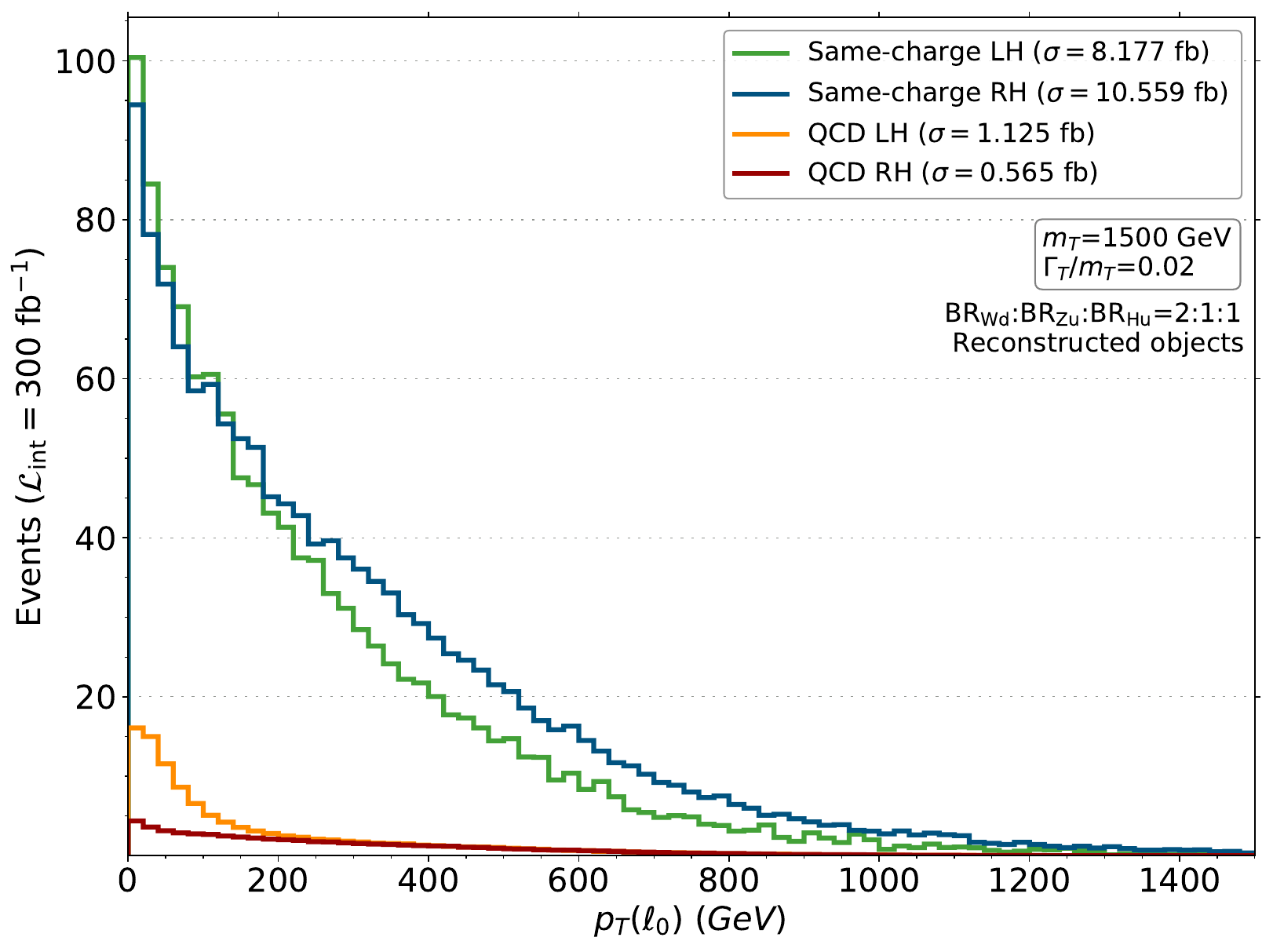}
\end{minipage}\hfill
\begin{minipage}{.49\textwidth}
\includegraphics[width=\textwidth]{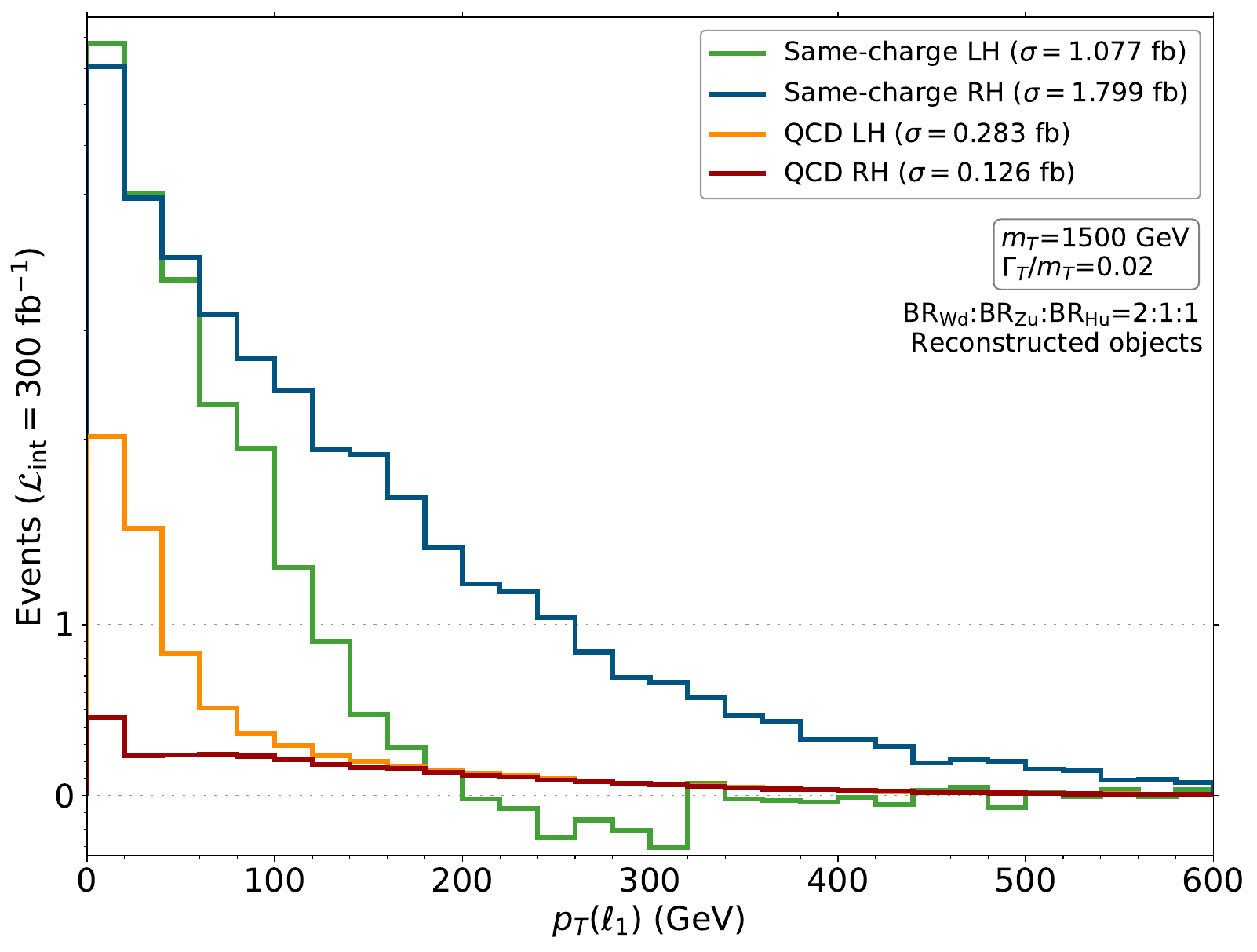}
\end{minipage}
\caption{\label{fig:SSvsQCDdist} Comparison between same-charge and QCD production and between LH and RH couplings for a singlet-like $T$ with $m_T=1.5$ TeV, $\Gamma_T/m_T=0.02$ and for different observables: from {top left} to {bottom right}, $m_{\rm eff}$ ({top}), $\Delta R(j_0,j_1)$, $p_T(j_0)$, $p_T(k_1)$, $p_T(\ell_0)$ and $p_T(\ell_1)$. The cross-sections associated to the distributions are effective to account for the fact that events may fall out of the range of the plots or may not contain the objects associated to the observables (such as $p_T$ of leptons in a fully-hadronic event).}
\end{figure}

A comparison between the same-charge and QCD productions and between is finally shown in \cref{fig:SSvsQCDdist}, for various observables, following the same procedure. 
Clearly, for this benchmark point, the difference in the cross-section is the main source of discrimination between the same-charge and QCD production processes. In some regions the QCD contribution is larger, but one should keep in mind that these distributions are obtained without imposing any cut to reduce the SM background. In the case of the $m_{\rm eff}$ distribution, such cuts would completely remove the low energy range, where the QCD distribution dominates with respect to the same-charge one in the LH coupling scenario. 
The differences between coupling chiralities are however large, especially for the same-charge production processes: in case of RH couplings the jets and leptons are produced with harder transverse momenta, which is also reflected in a larger number of events with high $m_{\rm eff}$.\\

\begin{figure}[h!]
\centering
\begin{minipage}{.49\textwidth}
\includegraphics[width=\textwidth]{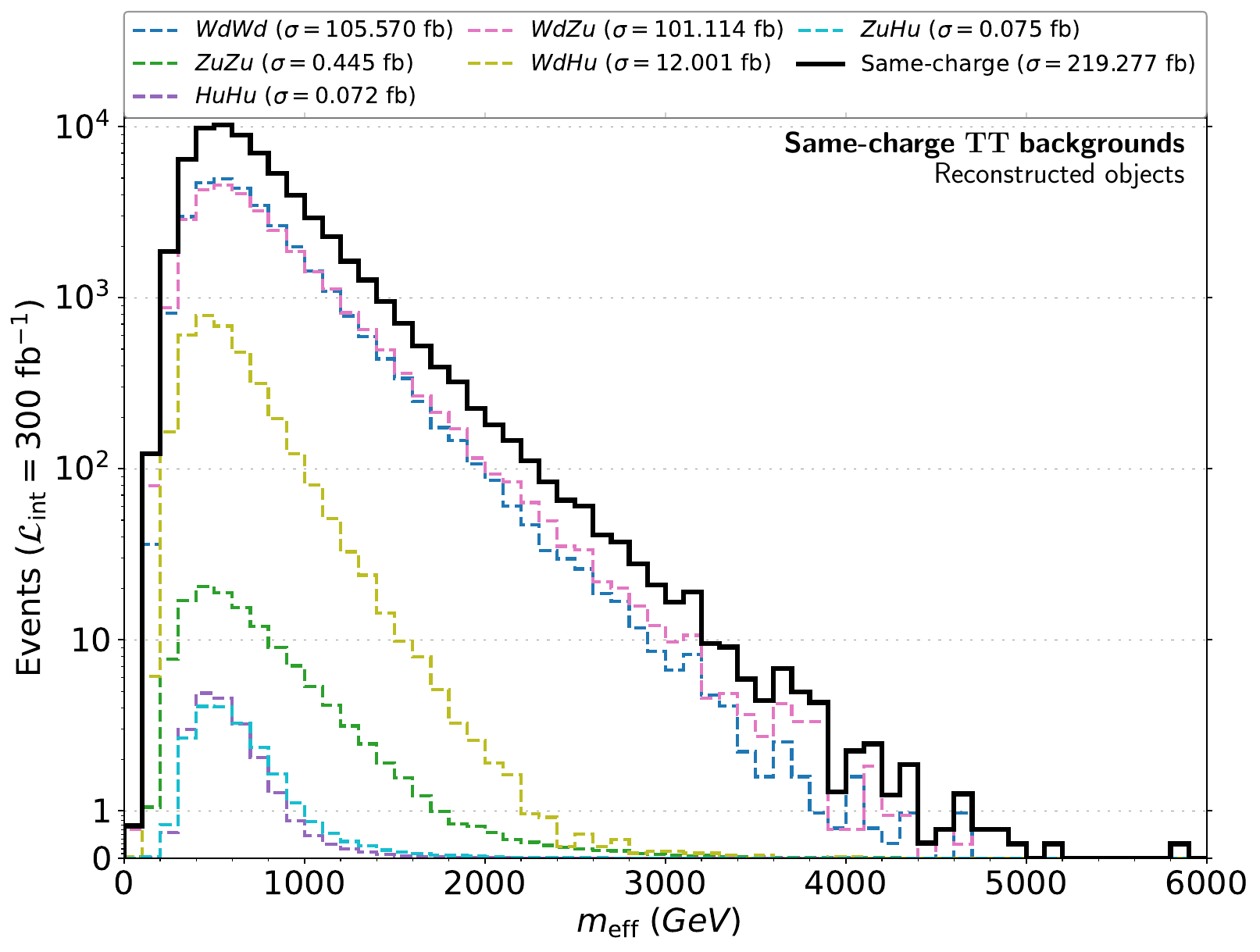}
\end{minipage}\hfill
\begin{minipage}{.49\textwidth}
\includegraphics[width=\textwidth]{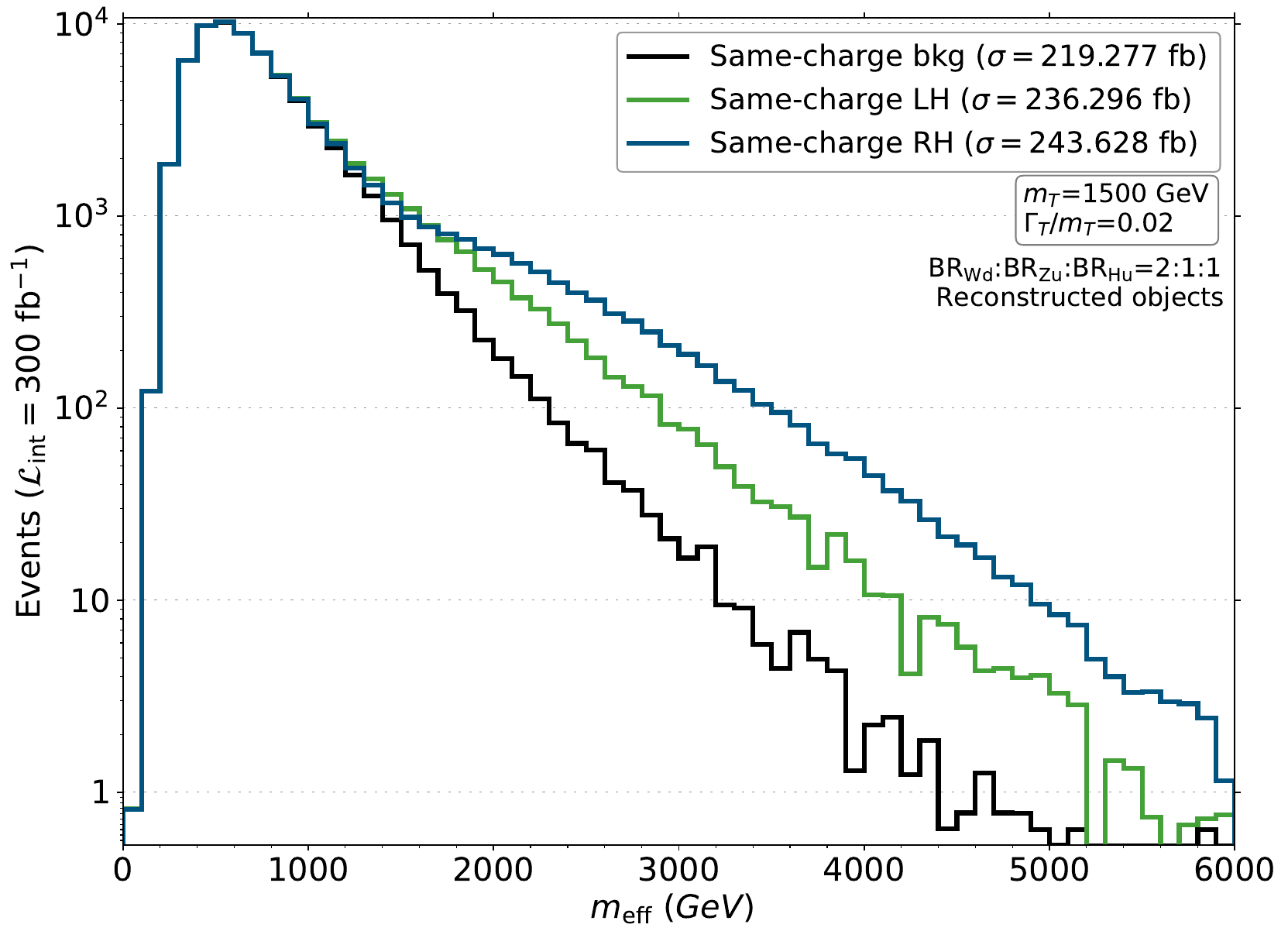}
\end{minipage}
\caption{\label{fig:SvsB} Breakdown of irreducible backgrounds for the same-charge process ({left}) and comparison between signal and background for the singlet-like $T$ with $m_T=1.5$ TeV and $\Gamma_T/m_T=0.02$ with LH or RH coupling chiralities.}
\end{figure}

We remind here that each contribution to the signal is unphysical if taken individually, and considering that interferences with the SM are also included, even the signal distributions shown so far are unphysical {\it per se} (bins with a negative number of events can indeed be seen in the $p_T(\ell_1)$ distributions). A {\it physical} cross-section (positive or at most null in case of maximal negative interference) and differential distributions can only be obtained when the signal is summed with the irreducible background. Considering the $m_{\rm eff}$ observable, the signal for the same-charge process in this example gives an overall excess of events, spread along the tail of the distribution, which can be relevant even with respect to the irreducible background, as shown in \cref{fig:SvsB}, even if a dedicated analysis including all backgrounds is in order to assess the observability of the signal and the potential to discriminate between coupling chiralities.

\section{Conclusions}

In summary, we have considered the pair production of a VLQ partner to the top quark of the SM, which we labelled as $T$, interacting solely with the first generation of SM quarks at the LHC. Due to its interactions with the up-quark, we have assessed the relevance of the same-charge $TT$ production process (induced by EW interactions) with respect to the opposite-charge $T\bar T$ one (induced by QCD interactions) usually considered in experimental searches. In comparison to the latter, the same-charge production channel has the known advantage of being enhanced by the up (valence) quark PDF and, if the coupling between the VLQ and SM particles is large enough, it can be of significant phenomenological relevance. 
This also implies that the VLQ width can be relatively large with respect to its mass, though, thus not allowing a consistent treatment within the NWA. We have therefore considered all the subprocesses leading to the same final state corresponding to the decay products of the $T$ without imposing its resonant propagation. Furthermore,  we have also accounted for all signal-background and signal-signal interference terms.

We have then presented various benchmark scenarios in a simplified model approach, where the $T$ VLQ production and decay phenomenology is described solely by its mass and couplings. As for the $T$ decay patterns, three are naturally possible and phenomenologically viable, which BRs have initially been chosen to be 100\%, into either of the possible  channels $Wd$, $Zu$ and $Hu$. Furthermore, we have also studied the case of two more theoretically driven decay patterns, wherein the BRs have asymptotic singlet-like (${\rm BR}_{T\to Wd}$:${\rm BR}_{T\to Zu}$:${\rm  BR}_{T\to Hu}$=2:1:1) and doublet-like (${\rm BR}_{T\to Wd}$:${\rm BR}_{T\to Zu}$:${\rm BR}_{T\to Hu}$=0:1:1) values. Indeed, we have tackled the case of two truly minimal VLQ extensions where $T$ is an EW singlet or part of a $(T~B)$ doublet, the latter under the further simplifying assumption that $B$  VLQs do not mix with the SM bottom quark. 

We have shown that, in all cases, the same-charge $TT$ production process has a strong potential to constrain scenarios with a $T$ interacting with SM first generation quarks. This process can in fact be competitive with opposite-charge $T\bar T$  production already for small values of the $T$ width (with respect to its mass) and can strongly dominate when the width is large enough. 

Constraints from low energy observables (such as APV data), however, have to be taken into account for a proper assessment of the validity of theoretical scenarios predicting a $T$ VLQ interacting with such light quarks when a complete model is considered. We have shown that such bounds can be competitive with the LHC ones, even if the continuous increment of integrated luminosity (and energy in the future) at the CERN machine is already allowing to probe  more regions of the allowed parameter space. We have then compared our results with previous literature, highlighting the differences in their determination. Finally, we have analysed, for some phenomenologically relevant masses and widths of the VLQ, the main features of the same-charge and QCD pair production processes which can allow for further observability and discrimination between signals.

In short, the PDF-enhanced same-charge pair production of $T$ VLQs which interact only with the SM first generation quarks can therefore be a very interesting channel to explore in future VLQ searches: it is very well  motivated theoretically and it allows for signatures that  have not been fully explored via dedicated searches, which we then advocate to take place.

\section*{Acknowledgements}

SM is supported in part through the NExT Institute and the STFC Consolidated Grant ST/X000583/1.
LP's work is supported by ICSC – Centro Nazionale di Ricerca in High Performance Computing, Big Data and Quantum Computing, funded by the European Union – NextGenerationEU.
LS is supported by the Natural Science Foundation of Henan Province under Grant No. 232300421217 and the China Scholarship Council under Grant No. 202208410277.

\section*{Dedication}

SM and LP dedicate this work to the dear memory of Prof. Elena Accomando, a friend and colleague.

\newpage
\appendix

\section{Breakdown of cross-sections for representative singlet-like $T$ points}
\label{app:breakdownxs}

\cref{tab:xssubprocesses} shows the breakdown of the contributions to the cross-sections for same-charge and QCD pair production of a singlet-like $T$ with mass $m_T=1$ TeV and $\Gamma_T/m_T=0.01$ and 0.1, and $m_T=1.5$ TeV with $\Gamma_T/m_T=0.02$, according to the deconstruction described in \cref{eq:sigmahats}. 
\begin{table}[h!]
\small
\centering
\begin{minipage}{.5\textwidth}
\centering
{\renewcommand{\arraystretch}{1}
\setlength{\tabcolsep}{0pt}
\begin{tabular}{cc|ccc}
\toprule
\midrule
\multicolumn{5}{c}{Same-charge $TT$ production}\\
\midrule
\midrule
\multirow{3}{*}{final state} & \multirow{3}{*}{process } & \multicolumn{3}{c}{cross-sections (fb)} \\
& & $\begin{array}{c}m_T=1~{\rm TeV}\\ {\Gamma_T\over m_T}=0.01\end{array}$ & $\begin{array}{c}1~{\rm TeV}\\0.1\end{array}$ & $\begin{array}{c}1.5~{\rm TeV}\\0.02\end{array}$ \\
\midrule
\multirow{5}{*}{$\begin{array}{c}W^+dW^+d\\[-1pt]+\\[2pt]W^-\bar d W^- \bar d\end{array}$} 
& $\sigma^Z$              &  3.213  &  455.138 &  1.629 \\
& $\sigma^H$              &  2.951  &  271.001 &  1.553 \\
& $\sigma^{ZB_{\rm int}}$ &  49.363 &  565.114 &  19.586 \\
& $\sigma^{HB_{\rm int}}$ &  5.906  &  58.615  &  3.525 \\
& $\sigma^{ZH_{\rm int}}$ & -6.018  & -591.958 & -3.104 \\
\midrule
\multirow{5}{*}{$\begin{array}{c}ZuZu\\[-1pt]+\\[2pt]Z\bar u Z \bar u\end{array}$} 
& $\sigma^Z$              &  0.770 &  73.364  &  0.391 \\
& $\sigma^H$              &  0.740 &  70.305  &  0.395 \\
& $\sigma^{ZB_{\rm int}}$ &  0.216 &  2.108   &  0.145 \\
& $\sigma^{HB_{\rm int}}$ &  1.844 &  17.880  &  1.350 \\
& $\sigma^{ZH_{\rm int}}$ & -1.488 & -132.399 & -0.769 \\
\midrule
\multirow{5}{*}{$\begin{array}{c}HuHu\\[-1pt]+\\[2pt]H\bar u H \bar u\end{array}$} 
& $\sigma^Z$              &  0.768 &  73.928  &  0.399 \\
& $\sigma^H$              &  0.742 &  70.311  &  0.387 \\
& $\sigma^{ZB_{\rm int}}$ &  1.345 &  14.516  &  0.776 \\
& $\sigma^{HB_{\rm int}}$ &  0.133 &  1.202   &  0.068 \\
& $\sigma^{ZH_{\rm int}}$ & -1.496 & -127.878 & -0.773 \\
\midrule
\multirow{5}{*}{$\begin{array}{c}W^+dZu\\[-1pt]+\\[2pt]W^-\bar d Z \bar u\end{array}$} 
& $\sigma^Z$              & 3.139     &  359.848   &  1.595 \\
& $\sigma^H$              & 2.953     &  272.907   &  1.566 \\
& $\sigma^{ZB_{\rm int}}$ & $\ll1$ ab &  $\ll1$ ab &  $\ll1$ ab \\
& $\sigma^{HB_{\rm int}}$ & $\ll1$ ab &  $\ll1$ ab &  $\ll1$ ab \\
& $\sigma^{ZH_{\rm int}}$ & 6.011     & -556.656   & -3.107 \\
\midrule
\multirow{5}{*}{$\begin{array}{c}W^+dHu\\[-1pt]+\\[2pt]W^-\bar d H \bar u\end{array}$} 
& $\sigma^Z$              &  3.138     &  358.112   &  1.611 \\
& $\sigma^H$              &  2.958     &  272.637   &  1.548 \\
& $\sigma^{ZB_{\rm int}}$ &  $\ll1$ ab &  $\ll1$ ab &  $\ll1$ ab \\
& $\sigma^{HB_{\rm int}}$ &  $\ll1$ ab &  $\ll1$ ab &  $\ll1$ ab \\
& $\sigma^{ZH_{\rm int}}$ & -6.017     & -554.633   & -3.107 \\
\midrule
\multirow{5}{*}{$\begin{array}{c}ZuHu\\[-1pt]+\\[2pt]Z\bar u H \bar u\end{array}$} 
& $\sigma^Z$              &  1.539 &  147.075 &  0.790 \\
& $\sigma^H$              &  1.481 &  140.372 &  0.782 \\
& $\sigma^{ZB_{\rm int}}$ & -1.198 & -12.206  & -1.008 \\
& $\sigma^{HB_{\rm int}}$ & -0.526 & -4.433   & -0.326 \\
& $\sigma^{ZH_{\rm int}}$ & -2.986 & -264.777 & -1.548 \\
\midrule
\multicolumn{2}{c|}{Total} & 57.439 & 979.493 & 24.351 \\
\midrule
\bottomrule
\end{tabular}
}
\end{minipage}\hfill
\begin{minipage}{.5\textwidth}
\centering
{\renewcommand{\arraystretch}{1}
\setlength{\tabcolsep}{0pt}
\begin{tabular}{cc|ccc}
\toprule
\midrule
\multicolumn{5}{c}{QCD $T\bar T$ production} \\
\midrule
\midrule
\multirow{3}{*}{final state } & \multirow{3}{*}{process } & \multicolumn{3}{c}{cross-sections (fb)} \\
& & $\begin{array}{c}m_T=1~{\rm TeV}\\ {\Gamma_T\over m_T}=0.01\end{array}$ & $\begin{array}{c}1~{\rm TeV}\\0.1\end{array}$ & $\begin{array}{c}1.5~{\rm TeV}\\0.02\end{array}$ \\
\midrule
\multirow{2}{*}{$W^+dW^-\bar d$} 
& $\sigma$           & 7.840 & 13.270 & 0.401 \\
& $\sigma^{\rm int}$ & 0.012 & 0.118  & 0.006 \\
\midrule
\multirow{2}{*}{$ZuZ\bar u$} 
& $\sigma$           & 2.052 & 4.167 & 0.116 \\
& $\sigma^{\rm int}$ & 0.009 & 0.087 & 0.003 \\
\midrule
\multirow{2}{*}{$HuH\bar u$} 
& $\sigma$           & 2.016     & 3.726     & 0.112 \\
& $\sigma^{\rm int}$ & $\ll1$ ab & $\ll1$ ab & $\ll1$ ab \\
\midrule
$\begin{array}{c} W^+dZ\bar u \\[-1pt]+\\[2pt] W^-\bar d Zu\end{array}$
& $\begin{array}{c}\sigma \\ \sigma^{\rm int} \end{array}$ & $\begin{array}{c} 8.010 \\ \ll1~{\rm ab}\end{array}$ & $\begin{array}{c} 13.602 \\ \ll1~{\rm ab}\end{array}$ & $\begin{array}{c} 0.426 \\ \ll1~{\rm ab} \end{array}$ \\
\midrule
$\begin{array}{c} W^+dH\bar u \\[-1pt]+\\[2pt] W^-\bar d Hu\end{array}$
& $\begin{array}{c}\sigma \\ \sigma^{\rm int} \end{array}$ & $\begin{array}{c} 8.009 \\ \ll1~{\rm ab}\end{array}$ & $\begin{array}{c} 20.448 \\ \ll1~{\rm ab}\end{array}$ & $\begin{array}{c} 0.491 \\ \ll1~{\rm ab} \end{array}$ \\
\midrule
\multirow{2}{*}{$ZuH\bar u$} 
& $\sigma$           &  8.460 &  49.833 &  0.400 \\
& $\sigma^{\rm int}$ & -0.678 & -6.739  & -0.310 \\
\midrule
\multicolumn{2}{c|}{Total} & 35.730 & 98.512 & 1.646 \\
\midrule
\bottomrule
\end{tabular}
}
\end{minipage}
\caption{\label{tab:xssubprocesses} Breakdown of contributions to the same-charge and QCD pair production cross-sections for a singlet-like $T$ (${\rm BR}_{T\to Wd}$:${\rm BR}_{T\to Zu}$:${\rm BR}_{T\to Hu}$=2:1:1) with RH couplings, mass $m_T=1$ TeV and two values of the total width/mass ratio, namely 0.01 and 0.1, and $m_T=1.5$ TeV with $\Gamma_T/m_T=0.02$.}
\end{table}

\newpage


\bibliography{literature.bib}
\bibliographystyle{JHEP}
\end{document}